\documentclass[english,11pt,oneside]{article}

\pdfoutput=1

\usepackage{amsmath}
\usepackage{amsfonts}
\usepackage{amssymb}
\usepackage{graphbox,graphicx, rotating}
\usepackage{epstopdf}
\usepackage{epsfig}
\usepackage{latexsym}
\usepackage{multirow}
\usepackage{color}
\usepackage[dvipsnames]{xcolor}
\usepackage{slashed}
\usepackage[top=2.6 cm, bottom=2.6 cm, left=2.8 cm, right=2.8 cm]{geometry}
\usepackage{amstext} 
\usepackage{array}  
\usepackage{hyperref}
\hypersetup{
colorlinks = true,
 linkcolor  = black,
 citecolor={blue}
}
\usepackage{bbold}

\usepackage{mcite}
\usepackage[utf8]{inputenc}
\usepackage{bm}
\usepackage{leftidx}
\usepackage[numbers,sort&compress]{natbib}

\usepackage{makecell}

\newcommand{\be}{\begin{equation}}
\newcommand{\ee}{\end{equation}}
\newcommand{\ba}{\begin{array}}
\newcommand{\ea}{\end{array}}
\newcommand{\eq}[1]{Eq.~(\ref{#1})}

\newcommand{\ab}{{\alpha\beta}}
\newcommand{\cd}{{\gamma\delta}}

\newcolumntype{L}{>{$}l<{$}}

\newcommand{\lv}[1]{\textcolor{purple}{#1}}

\begin{document}

\title{\bf Effective approach to lepton observables:\\ the seesaw case}

\author{Rupert Coy \& Michele Frigerio \\ 
{\small\emph{Laboratoire Charles Coulomb (L2C), University of Montpellier, CNRS, Montpellier, France}}\\
}
\date{}

\maketitle

\begin{abstract}
In the absence of direct evidence of new physics, any ultraviolet theory can be reduced to its specific set of low-energy effective operators. 
As a case study, we derive the effective field theory for the seesaw extension of the Standard Model, with sterile neutrinos of mass $M>m_W$.
We systematically compute all Wilson coefficients generated at one loop. 
Hence, it becomes straightforward to (i) identify the seesaw parameters compatible with the smallness of neutrino masses; (ii) compute precision lepton observables, which may be sensitive to scales as large as $M\sim 10^3$ TeV;
and (iii) establish sharp correlations among those observables. 
We find that the flavour-conserving Wilson coefficients set an upper bound on the flavour-violating ones. 
The low-energy limits on $\mu\to e$ and $\tau\to e,\mu$ transitions suppress flavour violation in $Z$ and Higgs decays, as well as electric  
dipole moments, far beyond the experimental reach. The precision measurements of $G_F$, $m_W$, and $Z$ partial decay widths set more stringent bounds than present and future limits on $\tau\to e,\mu$ transitions. 
We also present a general spurion analysis, to compare the seesaw with different models, thus assessing the discriminating potential of the effective approach.
%\begin{description}
%\end{description}
\end{abstract}

\newpage
\tableofcontents

\newpage

\section{Introduction}

Precision measurements in the lepton sector allow us to test the Standard Model (SM) to an exceptional depth. 
As no new particles have been detected so far, these measurements can be fully
described in terms of effective operators involving the SM fields only.
In most cases, the present bounds push the cutoff, $\Lambda$, of this effective field theory (EFT) well beyond the TeV scale.
Indeed, the dimension-five (dim-5) Weinberg operator \cite{Weinberg:1979sa} should have an extremely small coefficient, to account for the tininess of the neutrino masses $m_\nu$.
Among dim-6 operators, those inducing lepton flavour violation (LFV) and CP violation are also extremely constrained, and even the flavour-conserving ones are subject to a few stringent bounds, as we will see.

From a top-down perspective, any theory beyond the SM is defined by some set of heavy degrees of freedom with mass $\gtrsim \Lambda$, which would surely have an interesting phenomenology if they were directly produced, either in the early Universe or in the lab. 
Still, the low-energy predictions of the theory can be fully encoded in a set of Wilson coefficients (WCs) of the EFT valid below the cutoff $\Lambda$. 
We would like to argue that by computing such a set of WCs for a given theory, all unnecessary details of the underlying class of models are dismissed, thereby offering an optimal method to compare with other theories. 
This is not only a matter of principles: in the lepton sector, there is the concrete possibility of distinguishing between different ultraviolet completions, owing to several clean signatures that may become available in the near future.

To illustrate this programme, in this article we focus on the seesaw scenario \cite{Minkowski:1977sc,Yanagida:1979as,GellMann:1980vs,Mohapatra:1979ia}, which amounts to adding to the SM a set of gauge singlet chiral fermions, the sterile neutrinos. 
This model is often dubbed the type-I seesaw mechanism, to distinguish it from alternative possibilities for inducing non-zero neutrino masses. 
The Majorana mass scale of sterile neutrinos, $M$,  can span a very wide energy range, consequently sterile neutrinos may have a remarkable variety of phenomenological applications if they are directly produced, e.g. leptogenesis at very high scales 
\cite{Fukugita:1986hr,Davidson:2008bu}, 
collider searches at the TeV scale \cite{delAguila:2007qnc,Kersten:2007vk,Deppisch:2015qwa}, dark matter searches at the keV scale \cite{Dodelson:1993je,Shi:1998km,Shaposhnikov:2006xi}, and anomalies in oscillation experiments at the eV scale \cite{Dentler:2018sju}. 
Regrettably, leptogenesis is in general very difficult to test, and no clear evidence of sterile neutrino detection in the lab has emerged so far. 
Here we will rather focus on the indirect effects of heavy sterile neutrinos on the phenomenology of the SM leptons, assuming $M$ is larger than the energy of the experiment under consideration.
Even in this limit the phenomenology may be extremely rich. 

We will demonstrate that the seesaw EFT description elucidates the correlations among the various observables. 
Specifically, it will become straightforward to study the limiting case,
where the sterile neutrinos have masses not far above the electroweak scale as well as large Yukawa couplings, and nonetheless $m_\nu$ remains sufficiently small.
This case encompasses, in particular, the inverse-seesaw limit \cite{Wyler:1982dd,Mohapatra:1986bd,GonzalezGarcia:1988rw}, and it is the most interesting phenomenologically
as several lepton observables can be close to the experimental sensitivity. 
Studies of various aspects of the seesaw phenomenology using  EFT techniques have previously been performed in e.g. \cite{Broncano:2003fq,Cirigliano:2005ck,Antusch:2005gp,Abada:2007ux,deGouvea:2007qla,Gavela:2009cd}.

In section \ref{seesawEFT}, we derive the set of WCs induced by the seesaw. 
We present for the first time the full list that arises at one-loop leading-log order, after specialising the general formalism of Renormalisation Group Equations (RGEs) to the seesaw case. 
We also include an important case of one-loop matching.
The results hold for any set of seesaw parameters, within the regime of validity of the EFT.
Appendix \ref{sec:ass} presents a systematic diagonalisation procedure for the seesaw mass matrix, which is useful to complement the EFT approach. Appendix \ref{sec:RGEs} collects the complete list of relevant effective operators and RGEs.

In section \ref{sec:pheno}, we discuss the seesaw predictions for lepton observables in terms of the seesaw WCs. 
We discuss in turn Higgs and $Z$ boson decays, LFV in charged lepton transitions, dipole moments, and corrections to the Fermi constant.  
We improved upon existing analyses of several observables, both fixing errors in the literature and using more recent data. 
The EFT approach enables an immediate comparison of the different processes. Indeed, we identify some interesting limits and correlations, which were previously overlooked.
Our bounds on the seesaw parameters are summarised in tables \ref{rtrace} and \ref{remu}, as well as in figures \ref{emuPlot}, \ref{etauPlot} and \ref{mutauPlot}.

In section \ref{sec:spurions} we investigate to what extent the seesaw predictions follow from symmetry considerations only, and how the predictions may be different in other models. These questions are better addressed
with a spurion formalism, which clarifies the different possible patterns for flavour-symmetry breaking. We then summarise our main findings.

%%%%%%%%%%%%%%%%%%%%%%%%%%%%%%%%%%%%%%
\section{Effective field theory for the seesaw}\label{seesawEFT} 
%%%%%%%%%%%%%%%%%%%%%%%%%%%%%%%%%%%%%%

Let us consider extending the SM by $n_s$ sterile neutrinos  $N_R$, that is, chiral fermions singlet under the SM gauge interactions,
\begin{align}
\mathcal{L}_{\text{seesaw}} &= \mathcal{L}_{\text{SM}} + i \overline{N_R} \slashed\partial N_R  - \left( \frac 12 \overline{N_R} M {N_R}^c  + \overline{N_R} Y  {\tilde{H}^\dagger} l_L + h.c.\right)~.
\label{lag}
\end{align}
Here $M$ is the symmetric, $n_s\times n_s$ matrix of sterile neutrino Majorana masses and $Y$ is the $n_s\times 3$ matrix of neutrino Yukawa couplings.
Once the Higgs acquires a vacuum expectation value, $\langle H^0\rangle = v/\sqrt{2}\simeq 174$ GeV, the neutrino Dirac mass matrix is generated, $m\equiv Yv/\sqrt{2}$.
While the entries $m_{i\alpha}$ are bound to lie at or below the electroweak scale, the  
eigenvalues $M_1\le M_2 \le \dots \le M_{n_s}$ of the matrix $M$ can take any value between zero and the cutoff of the theory.
In the limit $m_{ia}\ll M_i$, the seesaw mechanism is realised \cite{Minkowski:1977sc,Yanagida:1979as,GellMann:1980vs,Mohapatra:1979ia},
and the $3\times 3$ Majorana mass matrix of light neutrinos takes the form $m_\nu \simeq - m^T M^{-1} m$. 
Note that the seesaw is operative for an extensive range of sterile masses, $m_\nu\sim 0.1$ eV $\ll M_i \lesssim v^2/m_\nu \sim 10^{15}$ GeV.

In the following we will derive the EFT below the scale $M_i$, which trades sterile neutrino interactions for higher dimensional operators involving only SM fields.
We will assume for definiteness sterile neutrinos heavier than the electroweak scale, $M_i > m_W$, but the same EFT techniques could be applied when (some of) the sterile neutrinos are lighter. 
At the scale $m_W$, the EFT involving SM multiplets will be matched to the EFT with broken electroweak symmetry.

Parts of this exercise have been presented in previous literature \cite{Weinberg:1979sa,Broncano:2002rw,Broncano:2004tz,Davidson:2018zuo,Jenkins:2013zja,Jenkins:2013wua,Alonso:2013hga}. 
Here we collect and generalise those results in a systematic fashion. In particular, we will apply the general RGEs for the SM effective operators to the seesaw case.
In addition, we include the one-loop matching of the dipole operators at the scales $M$ and $m_W$, which is necessary to correctly describe the lepton dipole transitions in the EFT language.

To fix the notation, we write the SM EFT Lagrangian as
\begin{align}
\mathcal{L}_{\text{SMEFT}} &= \mathcal{L}_{\text{SM}} + \frac{1}{\Lambda} \left(C^W Q_W + h.c.\right) + \frac{1}{\Lambda^2}\sum \limits_i \left(C^i  Q_i +h.c.\right) + {\cal O}\left(\frac{1}{\Lambda^3}\right) ~,
\label{EFTlag}
\end{align}
where $Q_W$ is the Weinberg operator, defined in table \ref{OpsM}, while $Q_i$ form a complete set of dim-6 operators, specifically we employ the Warsaw basis \cite{Grzadkowski:2010es}. 
The WCs, $C^W$ and $C^i$, are defined to be dimensionless, with $\Lambda$ the cutoff of the EFT, which may be identified for definiteness as the lightest sterile neutrino mass, $M_1$.
It is understood that the hermitian conjugate is not added to \eq{EFTlag} when an operator is self-hermitian, $Q_i=Q_i^\dag$. 
We will generally neglect operators with ${\rm dim}>6$, since most of the relevant observables are induced already by dim-5 and 6 operators and we are not interested in sub-leading corrections.
Exceptions will be discussed in due course.

%%%%%%%%%%%%%%%%%%%%%%%%%%%%%%%%%%%%%%
\subsection{Matching at the sterile neutrino mass scale $M$} \label{matchM}
%%%%%%%%%%%%%%%%%%%%%%%%%%%%%%%%%%%%%%

The seesaw Lagrangian (\ref{lag}) can be matched to the SM EFT (\ref{EFTlag}) by integrating out the sterile neutrinos at their mass scale $M_i$. 
At tree-level, it is sufficient to expand the $N_{R}$ equation of motion in inverse powers of $M$. This generates the dim-5 Weinberg operator \cite{Weinberg:1979sa} via the seesaw mechanism, and a linear combination of two dim-6 operators \cite{Broncano:2002rw}, 
\begin{align}
&\mathcal{L}_{M}^{\text{tree}} = \frac{1}{\Lambda}\left(C^W_{\ab} Q_{W,\alpha\beta} + h.c.\right) 
+ \frac{1}{\Lambda^2}   \left( C^{Hl(1)}_\ab Q_{Hl,\ab}^{(1)} + C^{Hl(3)}_\ab Q_{Hl,\ab}^{(3)} \right) ,
\label{EFTtreeM}
\end{align}
where the explicit form of the operators is provided in table \ref{OpsM}, and the WCs read
\begin{align}
& \frac{2}{\Lambda}C^W_\ab = (Y^T M^{-1} Y)_{\ab} = \frac{2}{\Lambda}\sum \limits_i C^{Wi}_\ab =  \sum \limits_i Y_{i\alpha} Y_{i\beta} M_i^{-1} ~, \label{CW}\\
& 
\frac{4}{\Lambda^2} C^{Hl(1)}_\ab =  - \frac{4}{\Lambda^2} C^{Hl(3)}_\ab = S_\ab \equiv (Y^\dagger M^{-1*} M^{-1} Y)_{\alpha\beta} = \sum \limits_i S^i_\ab = \sum \limits_i Y^*_{i\alpha} Y_{i \beta} M_i^{-2} \label{S}~,
\end{align}
where we conveniently introduced the hermitian matrix $S$, and the sums are defined in a basis with $M$ diagonal. 
The next order, matching onto dim-7 operators, is also known \cite{Elgaard-Clausen:2017xkq}.

\begin{table}[t]
\renewcommand{\arraystretch}{1.2}
\centering
\begin{tabular}{|c|c|} \hline
Name & Operator \\ \hline
$Q_{W,\alpha \beta}$ & $(\overline{l_{L\alpha}^c}  \tilde{H}^* ) (\tilde{H}^\dagger l_{L\beta} )$ \\ \hline
$Q_{Hl,\alpha \beta}^{(1)}$ & $(\overline{l_{L\alpha}} \gamma_\mu l_{L\beta})(H^\dagger i \overleftrightarrow{D^\mu} H)$ \\ \hline
$Q_{Hl,\alpha \beta}^{(3)}$ & $(\overline{l_{L\alpha}} \gamma_\mu \sigma^A l_{L\beta})(H^\dagger i \overleftrightarrow{D^\mu} \sigma^A H)$ \\ \hline 
$Q_{eB,\alpha \beta}$ & $(\overline{l_{L\alpha}} \sigma_{\mu \nu} e_{R\beta}) H B^{\mu \nu}$ \\ \hline
$Q_{eW,\alpha \beta}$ & $(\overline{l_{L\alpha}} \sigma_{\mu \nu} e_{R\beta}) \sigma^A H W^{A \mu \nu}$ \\ \hline
\end{tabular}
\caption{\small Operators generated by matching the seesaw at the sterile neutrino mass scale $M$.}
\label{OpsM}
\end{table}

\begin{figure}[t]
\begin{center}
\includegraphics[align=c,scale=0.55]{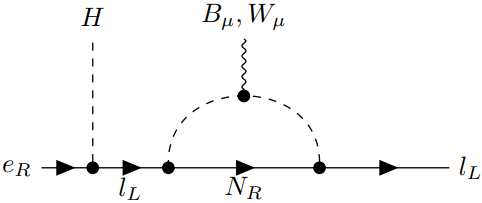}
\includegraphics[align=c,scale=0.5]{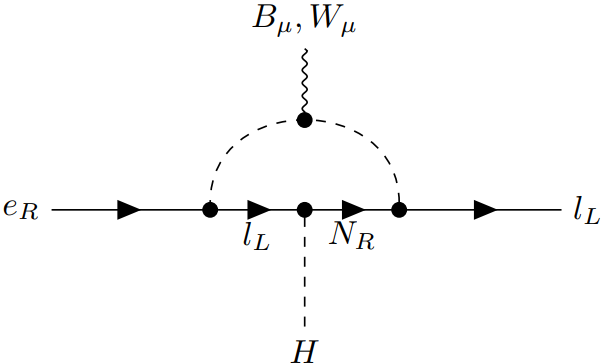}
\includegraphics[align=c,scale=0.45]{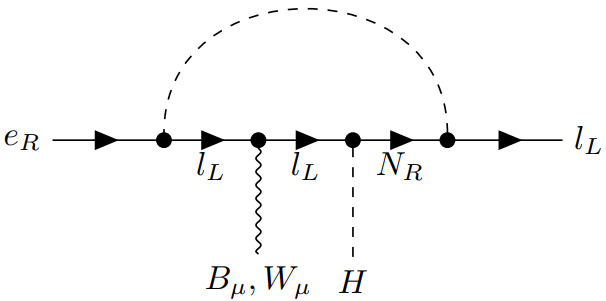}
%\vspace{-0.3cm}
\caption{\small One-loop diagrams that match onto the electroweak dipole operators $Q_{eB}$ and  $Q_{eW}$, at the mass scale $M$ of the sterile neutrino $N_R$.}
\label{DipoleM}
\end{center}
\end{figure}

Tree-level matching is not sufficient to describe the all-important dipole transitions, which are also induced by the sterile neutrinos. 
However, we remark that the EFT allows one to account for these effects consistently by matching the one-loop contribution of sterile neutrinos onto the electroweak dipole operators, which are also defined in table \ref{OpsM}. To this end, 
we computed the relevant diagrams, shown in Fig.~\ref{DipoleM}, which amounts to adding to the EFT Lagrangian the terms\footnote{
We performed the matching by computing the three diagrams of Fig.~\ref{DipoleM} on-shell. We would like to acknowledge that the final, correct result in \eq{EFTloopM} first appeared in Ref.~\cite{Zhang:2021tsq}, which performed the matching by employing an alternative, off-shell technique (Ref.~\cite{Zhang:2021tsq}  adopts the opposite convention for the sign in the covariant derivatives, and thus for the overall sign of the dipole operators). This prompted us to find a mistake in the previous version of our paper (though the outcome for the electromagnetic dipole WC remains the same).}
\be
\mathcal{L}_{M}^{\text{loop}} = 
\frac{1}{16 \pi^2} (S Y_e^\dagger)_\ab\left( -\frac{g_1}{24} Q_{eB,\ab} - \frac{5g_2}{24} Q_{eW,\ab} \right) + h.c.~, 
\label{EFTloopM}
\ee
where $g_1,g_2$ are the $U(1)_Y$ and $SU(2)_L$ gauge couplings, respectively, and $Y_e$ is the charged lepton Yukawa matrix, defined by \eq{SMLagrangian}. 
We will often replace it by its diagonal form, $(Y_e)_{\beta\gamma} = y_\beta \delta_{\beta\gamma}$ for $\beta=e,\mu,\tau$. 
Note the loop is finite and therefore does not induce any renormalisation-scale dependence.

The seesaw WCs can be concisely written in terms of the matrices  $Y$ and $M$ if one neglects the difference among the mass eigenvalues $M_i$. 
Strictly speaking, one should choose the basis where $M$ is diagonal and integrate out each mass eigenstate, $N_{Ri}$, at scale $\mu=M_i$,
that is, the seesaw parameters $Y_{i\alpha}$ and $M_i$ should be defined at that matching scale.
Still, if the matrices $Y$ and $M$ are defined at the largest seesaw scale, $M_{n_s}$, and their RG evolution to $M_1$ is neglected, one can show that the correction to the WCs is sub-leading.\footnote{
One should match the seesaw Lagrangian to an EFT with $n_s-1$ sterile neutrinos at the scale $\mu=M_{n_s}$, run this EFT down to $\mu=M_{n_s-1}$ and perform a new matching, and so on until the SM EFT is recovered
at $\mu=M_1$. 
This procedure introduces corrections to Eqs.~(\ref{EFTtreeM}) and (\ref{EFTloopM}) proportional to $\log M_i/M_j$, which are suppressed by an extra loop factor and extra couplings. 
Note also that below $M_j$, the EFT includes operators that combine SM fields and sterile neutrinos with mass $M_i<M_j$.
It has been shown \cite{Graesser:2007yj,delAguila:2008ir} that such an EFT contains dim-5 operators with two sterile neutrinos and dim-6 operators with one or more sterile states. 
One can check that, integrating out $N_{Ri}$ at scale $M_i$, these operators generate only SM operators with ${\rm dim} > 6$.
Of course, a detailed treatment of such intermediate scale effects may be relevant 
for a precision reconstruction of the seesaw parameters. See e.g. \cite{Antusch:2005gp} for the case of the Weinberg operator.}
On the other hand,  one cannot neglect the RG evolution of the WCs among the different scales $M_i$, because this affects the WCs at leading-logarithm order, as we will see below.

%

%%%%%%%%%%%%%%%%%%%%%%%%%%%%%%%%%%%%%%
\subsection{Running from $M$ to the electroweak scale $m_W$} 
\label{sec:RunningMmW}
%%%%%%%%%%%%%%%%%%%%%%%%%%%%%%%%%%%%%%

Let us discuss the evolution of the EFT Lagrangian from the sterile neutrino mass scale, $M_i$, to the electroweak scale, which we identify for definiteness as the $W$-boson mass, $m_W$.
The running of the only dim-5 operator, $Q_W$, is independent from dim-6 operators for dimensional reasons.
In contrast, the running of the dim-6 operators may receive
contributions from two insertions of $Q_W$. In equations,
\begin{equation}
\frac{dC^W}{d\log\mu} = \gamma_W C^W~,\qquad \frac{dC^i}{d\log\mu} = \gamma^i_j C^j + \gamma^i_{W} C^{W\dag} C^W~,
\label{RGEs}
\end{equation}
where 
$\gamma_W$, $\gamma^i_{j}$, $\gamma^i_{W}$ are the operator anomalous dimensions, and the appropriate contractions of flavour indices are understood.
The RGE for the Weinberg operator was calculated at one loop in {\cite{Chankowski:1993tx,Babu:1993qv,Antusch:2001ck}}.
A comprehensive compilation of the dim-6 anomalous dimensions in the SM EFT at one loop, $\gamma^i_{j}$, is provided in \cite{Jenkins:2013zja,Jenkins:2013wua,Alonso:2013hga}. 
We cross-checked a subset of these coefficients that are relevant for the seesaw. 
The mixing of the Weinberg operator (squared) into dim-6 operators, described by the coefficients $\gamma^i_{W}$, was calculated in \cite{Broncano:2004tz}, and one term has recently been corrected in \cite{Davidson:2018zuo}. 
We cross-checked this computation and agree with the latter result. 
The relevant operators and their complete one-loop RGEs are collected in appendix \ref{sec:RGEs}.

At the scale $M_i$, the only non-vanishing WCs are those
in Eqs.~(\ref{EFTtreeM}) and (\ref{EFTloopM}).
At lower scales, these operators source their own running and also mix into other operators, inducing additional non-zero WCs. 
To illustrate the result compactly, we take the
leading-logarithm approximation, and define 
\begin{align}
W_{\ab \gamma \delta} &\equiv \sum \limits_{i,j} \frac{4C^{Wi*}_{\alpha \gamma}C^{Wj}_{\beta \delta}}{\Lambda^2} \log\frac{\min(M_i,M_j)}{m_W} =   \sum \limits_{i,j} Y^*_{i\alpha} Y^*_{i\gamma} 
Y_{j\beta} Y_{j\delta} M_i^{-1} M_j^{-1}\log\frac{\min(M_i,M_j)}{m_W} ~,\label{Wdefn}
\end{align} %\\
\begin{align}
R_\ab &\equiv \sum \limits_i R^i_\ab = \sum \limits_i S_\ab^i \log\frac{M_i}{m_W}  = \sum \limits_i Y^*_{i\alpha} Y_{i\beta} M_i^{-2} \log\frac{M_i}{m_W}~,
\label{Rdefn}
\end{align}
which are generated, respectively, by two insertions of the WC in \eq{CW}, and one insertion of the WC in \eq{S}.
In the approximation where all logarithms are replaced by a common factor, $\log(M/m_W)$, then simply $W_{\ab\gamma\delta} \propto C^{W*}_{\alpha\gamma} C^W_{\beta\delta}$ and $R_\ab \propto S_\ab$.\footnote{
The approximation $\log(M_i/m_W)\simeq \log(M_1/m_W)$ for $R_\ab$ and $W_{\ab \gamma \delta}$ is tenable only if $M_i \simeq M_1$, or if the contribution of $N_{Ri}$ to the WCs is negligible w.r.t. the one of $N_{R1}$. 
For terms proportional to $R$, 
the latter condition reads $|Y_i^2/M_i^2| \log(M_i/M_1) \ll |Y_1^2/M_1^2| \log(M_1/m_W)$, where we have dropped flavour indices. 
For terms proportional to $W$, an analogous condition applies with $|Y_i^2/M_i^2|$ replaced by $|Y_i^4/M_i^2|$.} 

We find that the seesaw induces at $m_W$ five additional leptonic operators, which were vanishing at $M$, with WCs   
\begin{align}
&\frac{(C_{\alpha \beta}^{Hl(1)} + C_{\alpha \beta}^{Hl(3)})(m_W)}{\Lambda^2} \simeq - \frac{1}{16\pi^2}  \left[ 
\frac{g_1^2 + 17 g_2^2}{12}  R_{\alpha \beta} + \frac{g_1^2 - g_2^2}{6} \text{tr}(R) \delta_{\alpha \beta} - \frac 12 \sum_\gamma W_{\ab \gamma \gamma} \right] ~, \label{HlmW}\\
&\frac{C^{He}_{\alpha \beta}(m_W)}{\Lambda^2}  \simeq \frac{1}{16\pi^2} \left[ 
\frac{1}{2} y_\alpha R_{\alpha \beta} y_\beta - \frac{1}{3} g_1^2 \text{tr}(R) \delta_{\alpha \beta} 
\right] ~, \label{HemW} \\
&\frac{C^{ll}_{\alpha \beta \gamma \delta} (m_W)}{\Lambda^2}  \simeq \frac{1}{16\pi^2} \left[ 
\frac{g_1^2 - g_2^2 }{24} (R_{\alpha \beta} \delta_{\gamma \delta} + \delta_{\ab} R_{\gamma \delta}) 
+ \frac{g_2^2}{12} \left( R_{\alpha \delta} \delta_{\gamma\beta} + \delta_{\alpha\delta} R_{\gamma \beta}  \right) + \frac 12 W_{\ab \gamma \delta} \right] , \label{llmW}\\
&\frac{C^{le}_{\alpha \beta \gamma \delta} (m_W)}{\Lambda^2} \simeq \frac{1}{16 \pi^2} \frac{g_1^2}{6} R_{\alpha \beta} \delta_\cd  ~, \label{lemW} \\
&\frac{C^{eH}_{\alpha \beta} (m_W)}{\Lambda^2}  \simeq \frac{1}{16\pi^2} \left[ 
2 \lambda R_{\alpha \beta} + \frac{1}{3} g_2^2 \text{tr}(R) \delta_{\alpha \beta} - \frac 32 \sum_\gamma W_{\alpha \beta \gamma \gamma} + 2 \sum_{\gamma,\delta}W_{\gamma \gamma\delta\delta} \ \delta_{\alpha \beta} 
\right]  y_\beta ~, \label{eHmW}
\end{align}
where the indices $\alpha , \beta, \gamma , \delta$ run over $e,\mu,\tau$, and $\lambda$ is the quartic Higgs coupling defined in \eq{SMLagrangian}. 
Note that we simplified the full expression of the anomalous dimensions, found in appendix \ref{sec:RGEs}, by neglecting 
the charged lepton Yukawa couplings $y_\alpha$ relative to the other relevant SM couplings, as they are much smaller (for 
example $y_\tau \ll g_{1,2}$ even at very high scales \cite{Buttazzo:2013uya}). 
The WCs which are already non-zero at scale $M$ 
receive similar corrections, which are loop-suppressed with respect to their leading-order value: we will neglect those.

The RG evolution also induces two-lepton, two-quark ($2q2\ell$) operators, which are relevant to estimate the $\mu\to e$ conversion rate in nuclei (see section \ref{mutoeC}), 
\begin{align}
\frac{C_{\alpha \beta xy}^{lq(1)}(m_W)}{\Lambda^2} &\simeq - \frac{1}{16 \pi^2} R_{\alpha \beta} \left[ \frac{1}{4} (Y_u^\dagger Y_u - Y_d^\dagger Y_d )_{xy} + \frac{g_1^2}{36} \delta_{xy} \right]  ~, \label{lq1mW}\\
\frac{C_{\alpha \beta xy}^{lq(3)}(m_W)}{\Lambda^2} &\simeq - \frac{1}{16 \pi^2} R_{\alpha \beta} \left[ \frac{1}{4} (Y_u^\dagger Y_u + Y_d^\dagger Y_d )_{xy} - \frac{g_2^2}{12} \delta_{xy} \right]  ~,
\label{lq3mW} \\
%\end{align}
%\begin{align}
\frac{C_{\alpha \beta xy}^{lu}(m_W)}{\Lambda^2} &\simeq -  \frac{1}{16 \pi^2} R_{\alpha \beta} \left[-\frac{1}{2} (Y_u Y_u^\dagger)_{xy} + \frac{g_1^2}{9} \delta_{xy} \right]  ~, \label{lumW} \\
\frac{C_{\alpha \beta xy}^{ld}(m_W)}{\Lambda^2} &\simeq - \frac{1}{16 \pi^2} R_{\alpha \beta} \left[ \frac{1}{2} (Y_d Y_d^\dagger)_{xy} - \frac{g_1^2}{18} \delta_{xy} \right]  ~, \label{ldmW}
\end{align}
where $x,y$ are quark flavour indices, the quark Yukawa couplings are defined in \eq{SMLagrangian}, and we adopted the same simplifications as above. 
In addition, the seesaw tree-level operators mix into two operators which modify the Higgs boson kinetic term and therefore affect its  couplings (see section \ref{hdecay}), 
\begin{align}
&\frac{C^{HD}(m_W)}{\Lambda^2} \simeq \frac{1}{16 \pi^2} \left[ \frac{2}{3} g_1^2 \text{tr}(R) + 4 \sum_{\gamma,\delta}W_{\gamma \gamma\delta\delta} \right] ~, \label{HDmW} \\
&\frac{C^{H\square} (m_W)}{\Lambda^2} \simeq \frac{1}{16\pi^2} \left[ \left( \frac{1}{2} g_2^2 + \frac{1}{6} g_1^2 \right) \text{tr}(R) + \frac 12  \sum_{\gamma,\delta}W_{\gamma \gamma\delta\delta} \right]  ~. \label{HsquaremW}
\end{align}
The additional operators induced by the seesaw and their RGEs, listed for completeness in appendix \ref{sec:RGEs}, have no impact on the lepton observables that we shall analyse.

Some comments are in order on the quality of our approximations.
The leading-log contributions to the WCs are expected to dominate over one-loop finite parts as long as $\log(M_i/m_W)$ is significantly larger than one. 
On the other hand, dim-6 operators have observable consequences for $M_i$ not too far above $m_W$. When the logarithm becomes of order one, the leading-log term still gives the correct order of magnitude, barring possible cancellations. 
This issue will be addressed for specific observables in section \ref{sec:pheno}.
We will neglect systematically two-loop corrections. In particular, the running of the dipole operators, $Q_{eB,eW}$, and their mixing into other operators are two-loop suppressed, as the dipole WCs are themselves already one-loop suppressed.

Finally, we have treated the right-hand side of the RGEs in \eq{RGEs} as a constant. Of course, it is a function of SM couplings and WCs, which run at one-loop. 
This induces two-loop-order corrections to the WCs at $m_W$, which may be sizeable if $\log(M/m_W)$ is large and the couplings run quickly. 
A recent analysis of this effect can be found in \cite{Buras:2018gto}. 
In the seesaw, we find that such corrections are typically of order $\sim 10\%$, as illustrated at the end of appendix \ref{sec:RGEs}, and we will neglect them.
When precision is needed, one can perform an RGE-improved computation to account for these corrections, by integrating numerically the system of RGEs provided in appendix \ref{sec:RGEs} together with the RGEs for the SM parameters, provided for instance in \cite{Buttazzo:2013uya}.

%%%%%%%%%%%%%%%%%%%%%%%%%%%%%%%%%%%%%%
\subsection{Matching at $m_W$ and running to the charged lepton mass scale $m_\alpha$}
\label{sec:matchmw}
%%%%%%%%%%%%%%%%%%%%%%%%%%%%%%%%%%%%%%

At the electroweak scale, the SM states with mass $\mathcal{O}(m_W)$ must be 
by integrated out, namely the Higgs, $W$ and $Z$ bosons (and the top quark, which plays no role for the lepton observables). 
One is left with an EFT for massive leptons and quarks, with gauge symmetry $SU(3)_{QCD}\times U(1)_{QED}$.

A basis for the operators of such an EFT has been defined in \cite{Jenkins:2017jig}, and the matching of the SM EFT WCs onto this basis is provided in appendix C of that reference, up to
terms that are Yukawa-coupling suppressed. 
As we are interested in charged LFV processes and dipole moments, we need only consider the four-fermion operators involving charged leptons,
and the electromagnetic dipole operator. 
In the low-energy EFT, four-fermion operators are defined as
\begin{equation}
\mathcal{O}_{\psi \chi,\ab \cd}^{A,XY} = (\overline{\psi_\alpha} \Gamma_A P_X \psi_\beta) (\overline{\chi_\gamma} \Gamma_A P_Y \chi_\delta)~,
\end{equation}
where $\psi, \chi = \nu,e,u,d$ are mass eigenstates, $X = L,R$ with $P_{L,R}$ the chiral projectors, and $A = S,V,T$ with $\Gamma_S = \mathbb{1}$, $\Gamma_V = \gamma_\mu$ and $\Gamma_T = \sigma_{\mu \nu}$.
We restrict ourselves to vector-vector operators, because scalar-scalar operators are relatively suppressed by two powers of Yukawa couplings and 
therefore have negligible effects on the observables of interest. Four-fermion tensor operators are not generated in the seesaw at leading-log order.

Let us begin with operators with four charged leptons, which in the seesaw receive contributions from Eqs.~(\ref{HlmW})\textendash (\ref{lemW}). 
The matching conditions at $\mu = m_W$ read 
\begin{align}
C_{ee,\alpha \beta \gamma \delta}^{V,LL} &= C_{\alpha \beta \gamma \delta}^{ll} + \frac{1}{2} \left( - 1 + 2 s_w^2 \right) \Bigg[ \left( C_{\alpha \beta}^{Hl(1)} + C_{\alpha \beta}^{Hl(3)} \right) \delta_{\gamma \delta} + \delta_{\ab} \left( C_{\gamma \delta}^{Hl(1)} + C_{\gamma \delta}^{Hl(3)} \right) \Bigg]  ~,
\label{matcheell} \\
C_{ee,\alpha \beta \gamma \delta}^{V,LR} &= C_{\alpha \beta \gamma \delta}^{le} + 2 s_w^2 \left( C_{\alpha \beta}^{Hl(1)} + C_{\alpha \beta}^{Hl(3)} \right) \delta_\cd + \left(-1+2 s_w^2 \right) \delta_\ab C_{\gamma \delta}^{He} \label{matcheelr} ~,\\
C_{ee,\alpha \beta \gamma \delta}^{V,RR} &= s_w^2 \left( C_{\alpha \beta}^{He} \delta_{\gamma \delta} + \delta_{\alpha \beta} C_{\gamma \delta}^{He} \right) \label{matcheerr} ~,
\end{align}
where $s_w$ is the sinus of the weak mixing angle. 
Note that these equations do not involve $[C^{Hl(1)} - C^{Hl(3)}]$, therefore all these WCs are loop suppressed.

The operators with two charged leptons and two quarks, relevant for $\mu\to e$ conversion in nuclei, match according to 
\begin{align}
C_{eu,\alpha \beta xy}^{V,LL} &= V_{x w} V^*_{yz} \left(C_{\alpha \beta wz}^{lq(1)} - C_{\alpha \beta wz}^{lq(3)} \right) + \left( 1 - \frac{4}{3} s_w^2 \right) \left( C_{\alpha \beta}^{Hl(1)} + C_{\alpha \beta}^{Hl(3)} \right)\delta_{xy} \label{matcheull} ~,\\
C_{eu,\alpha \beta xy}^{V,LR} &= C_{\alpha \beta xy}^{lu} - \frac{4}{3} s_w^2 \left( C_{\alpha \beta}^{Hl(1)} + C_{\alpha \beta}^{Hl(3)} \right) \delta_{xy} \label{matcheulr} ~,\\
C_{ed,\alpha \beta xy}^{V,LL} &= C_{\alpha \beta xy}^{lq(1)} + C_{\alpha \beta xy}^{lq(3)} + \left( - 1 + \frac{2}{3} s_w^2 \right) \left( C_{\alpha \beta}^{Hl(1)} + C_{\alpha \beta}^{Hl(3)} \right) \delta_{xy} \label{matchedll} ~,\\
C_{ed,\alpha \beta xy}^{V,LR} &= C_{\alpha \beta xy}^{ld} + \frac{2}{3} s_w^2 \left( C_{\alpha \beta}^{Hl(1)} + C_{\alpha \beta}^{Hl(3)} \right) \delta_{xy} \label{matchedlr}  ~,
\end{align}
where $x,y,w,z$ are quark mass eigenstate indices, and in Eqs.~(\ref{lq1mW})-(\ref{ldmW})
we chose a basis where $Y_d=diag(y_d,y_s,y_b)$ and $Y_u = diag(y_u, y_c, y_t)V$, with $V$ the CKM matrix. 
Even this set of WCs does not depend on the combination $[C^{(1)}_{Hl}-C^{(3)}_{Hl}]$, so they all vanish at tree-level.

We will generally ignore operators involving neutrinos, which are typically less constrained (for a detailed discussion and special cases see e.g. \cite{Bergmann:1998ft,Bergmann:1999pk,Gavela:2008ra,Farzan:2017xzy}). 
One exception is the operator ${\cal O}^{V,LL}_{\nu e}$, which corrects
$\mu$ and $\tau$ beta-decays and is induced at tree-level. Its matching reads
\begin{equation}
C^{V,LL}_{\nu e,\ab \gamma \delta} \simeq - 2 \left( C^{Hl(3)}_{\alpha \delta} \delta_{\gamma\beta } + \delta_{\alpha \delta} C^{Hl(3)}_{\gamma \beta} \right) ~,
\label{VLLnu}
\end{equation}
where we neglected subdominant loop-level contributions. 
This is relevant to test the universality of the Fermi coupling, see section \ref{univL}.

\begin{figure}[t!]
\begin{center}
\vspace{-0.5cm}
\includegraphics[width=0.85\textwidth]{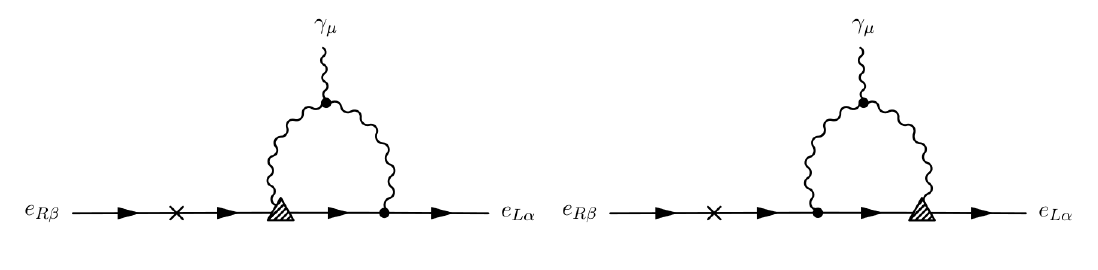}
\vspace{-0.6cm}
\caption{\small 
Relevant diagrams for the one-loop matching of the operator $Q_{Hl}^{(3)}$, in the seesaw EFT above $m_W$, 
onto the operator $\mathcal{O}_{e\gamma}$, in the EFT below $m_W$. 
The shaded triangle vertex stands for the $Q_{Hl}^{(3)}$ component $g_2 \langle H^0 H^0 \rangle  \overline{\nu} \slashed{W}^+ P_L e$. The wavy (arrow) lines in the loops
stand for $W$ bosons (active neutrinos).}
\label{chlMW}
\end{center}
\end{figure}

Finally, the electromagnetic dipole operator, $\mathcal{O}_{e\gamma,\ab}\equiv \overline{e_{L\alpha}}\sigma_{\mu\nu}e_{R\beta}F^{\mu\nu}v/\sqrt{2}$, 
receives contributions both from $C^{eB,eW}$ and from one-loop matching at the electroweak scale, 
\begin{align}
C_{e\gamma,\alpha \beta} &= c_w  C_{\alpha \beta}^{eB} - s_w C^{eW}_{\alpha \beta} + C_{e\gamma,\alpha \beta}^{\rm EW-h} + C_{e\gamma,\alpha \beta}^{\rm EW-l} ~.
\label{matchdipole} 
\end{align}
The term $C_{e\gamma}^{\rm EW-h}$ is generated by two diagrams, shown in Fig.~\ref{chlMW}, which correspond to the one-loop matching of $Q_{Hl}^{(3)}$ onto $Q_{e\gamma}$. 
They are finite (no renormalisation-scale dependence) and induce a contribution 
\begin{equation}
\frac{1}{\Lambda^2} C_{e\gamma,\ab}^{\rm EW-h} = \frac{5e (C^{Hl(3)} Y_e^\dagger)_\ab}{48\pi^2 \Lambda^2} ~.
\end{equation}
This is comparable to the contribution of one-loop matching at the sterile-neutrino mass scale $M$, given in Eq.~(\ref{EFTloopM}). Adding them we obtain
\begin{equation}
\frac{C^{\rm h}_{e\gamma,\alpha \beta}}{\Lambda^2} \equiv \frac{c_w  C_{\alpha \beta}^{eB} - s_w C^{eW}_{\alpha \beta} + C_{e\gamma,\alpha \beta}^{\rm EW-h}}{\Lambda^2} 
= - \frac{e (SY_e^\dagger)_\ab}{64\pi^2}~,
\label{heavyD}\end{equation}
which corresponds to the total contribution of the heavy neutrino mass eigenstates to the electromagnetic dipole.
We checked that our result matches (and generalises) the calculation of the heavy-neutrino contribution to $\ell_\alpha \to \ell_\beta \gamma$ in \cite{Cheng:1980tp,Altarelli:1977zq}, up to corrections suppressed by additional powers of the active-sterile mixing, which correspond to EFT operators with dim~$>6$.

\begin{figure}[t!]
\begin{center}
\vspace{-0.5cm}
\includegraphics[width=0.85\textwidth]{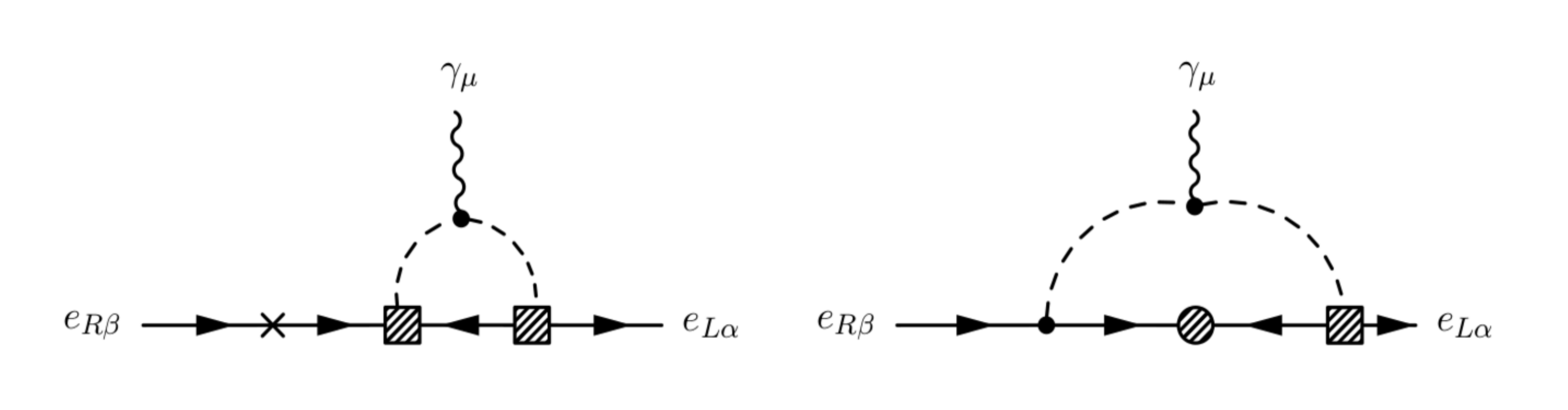}\\
\vspace{-0.5cm}
\includegraphics[width=0.85\textwidth]{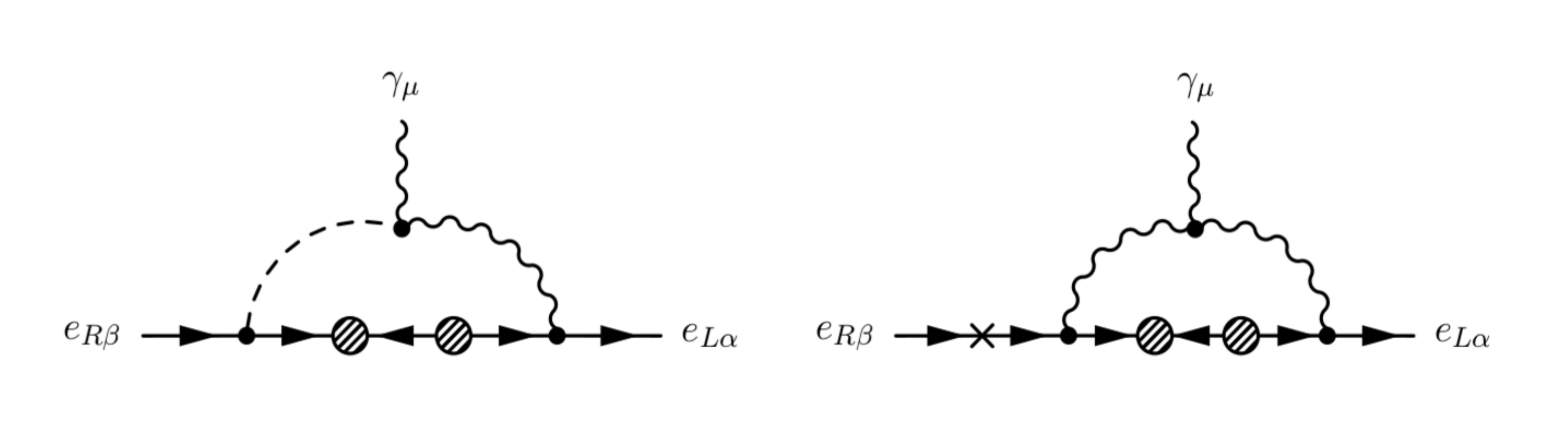}
\vspace{-0.7cm}
\caption{\small Relevant diagrams, in the 't Hooft-Feynman gauge, for the one-loop matching of the operator $Q_W$, in the seesaw EFT above $m_W$, 
onto the operator ${\cal O}_{e\gamma}$, in the EFT below $m_W$. 
The shaded square (circle) vertex stands for the Weinberg-operator component $\nu e^- H^+\langle H^0\rangle $ ($\nu \nu \langle H^0 H^0\rangle$). The wavy/dashed/arrow 
lines in the loops
stand for $W$ bosons / $H^+$ Goldstone bosons / active neutrinos.}
\label{DipoleMW}
\end{center}
\end{figure}

The term $C_{e\gamma}^{\rm EW-l}$ corresponds to the contribution of light neutrino mass eigenstates. In our EFT approach we find that, in the 't Hooft-Feynman gauge, it is generated 
by the four diagrams displayed in Fig.~\ref{DipoleMW}, which involve two insertions of the Weinberg operator $Q_W$.
The diagrams are finite and the result of the matching is
\begin{align}
\frac{C_{e\gamma,\alpha \beta}^{\rm EW-l}}{\Lambda^2} \frac{v}{\sqrt{2}} &= - \frac{e v^2 (C^{W\dag} C^W Y_e^\dag)_\ab }{64 \pi^2 m_W^2 \Lambda^2} \frac{v}{\sqrt{2}} = - \frac{e^3 U_{\alpha i} U^*_{\beta i} m_i^2 m_\beta }{256 \pi^2 s_w^2 m_W^4} ~,
\label{cegMatchMw}
\end{align}
where in the last equality we used $C^W v^2/\Lambda \simeq m_\nu = U^* diag(m_1,m_2,m_3) U^\dag$.
Higher order corrections to the neutrino masses and to the PMNS matrix, $U$, are discussed in appendix \ref{sec:ass}. 
The EFT result of \eq{cegMatchMw} allows us to reproduce, in particular, the result of the classical computation of the $\mu \to e \gamma$ decay width, in the SM augmented with light, massive neutrinos \cite{Petcov:1976ff,Marciano:1977wx,Lee:1977tib,Cheng:1976uq} (presented in full detail e.g. in \cite{Cheng:1985bj}).

To study charged lepton observables, such as LFV decays or dipole moments, it is necessary, in principle, to include the RG evolution of the WCs from the electroweak scale to the mass scale of the heaviest lepton involved in the process, $m_\alpha$. 
However, one can convince oneself that in the seesaw this running has no significant effects.
The only interactions below the electroweak scale are QED and QCD, both of which are flavour-conserving, therefore the structure of flavour-violation is fixed by the matching at $m_W$ and it does not change at lower scales.\footnote{
If the analysis is extended to dim-8 operators, two four-fermion operators can be combined in a `fish' diagram, which may induce additional flavour violation. We neglect such sub-leading effects here.}
In addition, the potentially large QCD corrections vanish at leading order for all the WCs in Eqs.~\eqref{matcheell}-\eqref{matchdipole}. 
The quark current in Eqs.~\eqref{matcheull}-\eqref{matchedlr} does not renormalise, up to terms chirally-suppressed by quark masses.
Therefore the only effect of the RG evolution from $m_W$ to $m_\alpha$ amounts to small QED corrections of order $\alpha/(4\pi)\,\log(m_W/m_\alpha) \lesssim 1\%$, which can be safely neglected in our analysis. 
In models other than the seesaw, the QED and QCD evolution of WCs from $m_W$ to the charged lepton mass scale can be relevant, see e.g. the analysis of flavour-violating observables in \cite{Crivellin:2017rmk}.

%%%%%%%%%%%%%%%%%%%%%%%%%%%%%%%%%%%%%%
\section{Phenomenological implications} 
\label{sec:pheno}
%%%%%%%%%%%%%%%%%%%%%%%%%%%%%%%%%%%%%%
%
We aim to employ the seesaw EFT, derived in section \ref{seesawEFT}, to compute various leptonic observables at leading order. 
Let us comment on the size of the next-to-leading corrections that we neglect. 
For observables induced at tree-level, the error is relatively suppressed by a loop factor times a log.
Similarly, those induced by finite one-loop diagrams (the dipoles) receive corrections suppressed by an extra loop times a log.
In contrast, observables generated by the one-loop RG evolution, and so are at leading-log order, receive finite one-loop corrections
whose relative size is $\sim\log^{-1}(M/m_W)$. Therefore, when $M$ is close to the TeV scale, the error may become large, as we will  discuss in some specific
cases below.

It is instructive to begin with some general considerations on the relative size of $C^W_\ab$ and $S_\ab$, defined in Eqs.~\eqref{CW} and \eqref{S}. The former is bound by the smallness of neutrino masses,
$|(m_\nu)_\ab| \simeq |C^W_\ab|v^2/(2\Lambda) \lesssim 0.1$ eV. 
Consequently the contributions to dim-six WCs proportional to $W$, defined in \eq{Wdefn}, are too small to have any observable consequence in all processes other than oscillations.
Thus, in the following we will systematically neglect them relative to the terms proportional to $R$, defined in \eq{Rdefn}.

A complementary question is whether the absolute size of $S_\ab$ is constrained by the smallness of neutrino masses, that is to say, whether the limit $C^W_\ab\to 0$ imposes some restriction on the size of $S_\ab$. 
This limit can be justified by an approximate lepton number symmetry, $U(1)_L$, which is realised e.g. in the inverse seesaw \cite{Wyler:1982dd,Mohapatra:1986bd,GonzalezGarcia:1988rw}. 
For phenomenological purposes one can even be more general, and consider the limit $C^W_\ab\to 0$  as a tuning of the seesaw parameters, which may or may not be justified by an underlying symmetry (see e.g. the discussion in  \cite{Kersten:2007vk}). 
Thus, we solve the system of equations
\be
\sum_{i=1}^{n_s} Y_{i\alpha} Y_{i\beta} M_i^{-1} = 0~,\qquad \alpha,\beta=e,\mu,\tau~.
\label{Wto0}\ee
For $n_s=2$, the general solution for the neutrino Yukawa coupling matrix reads
\be
Y= \begin{pmatrix} 1 \\ \pm i \sqrt{\dfrac{M_2}{M_1}}  \end{pmatrix} \begin{pmatrix}  \lambda_e ~~ \lambda_\mu ~~ \lambda_\tau \end{pmatrix} %\quad
\Rightarrow%\quad 
S_\ab = \frac{\lambda_\alpha \lambda_\beta}{M_1^2} \left(1+\frac{M_1}{M_2} \right), 
\label{rab2}
\ee
where $\lambda_{e,\mu,\tau}$ are arbitrary numbers that can be taken real and positive, by choosing the phases of $l_{Le,\mu,\tau}$. 
For $n_s=3$, the general solution of \eq{Wto0} reads
\begin{align}
&Y= \begin{pmatrix} 1 \\ z \sqrt{\dfrac{M_2}{M_1}} \\ \pm i \sqrt{1+z^2} \sqrt{\dfrac{M_3}{M_1}}  \end{pmatrix} \begin{pmatrix}  \lambda_e ~~ \lambda_\mu ~~ \lambda_\tau  \end{pmatrix} ~~\Rightarrow~~
S_\ab = \frac{\lambda_\alpha \lambda_\beta}{M_1^2} \left(1+|z|^2\frac{M_1}{M_2}+|1+z^2|\frac{M_1}{M_3} \right), 
\label{rab3}
\end{align}
where $z$ is an arbitrary complex number. 
Thus Eqs. \eqref{rab2} and \eqref{rab3} give the general analytic result for the dim-6 Wilson coefficients, $S_{\ab}$, in the limit of a vanishing dim-5 operator, for $n_s = 2,3$ respectively. 
To our knowledge, such expressions were not previously stated in the literature.
The contributions of the two (three) sterile neutrinos  to $S$ add constructively, so no cancellation is possible. 
Since the general solution for $Y$ is factorised (column times row), one finds that the matrix $S$ is factorised as well, $S_\ab \propto \lambda_\alpha \lambda_\beta$,
so only three entries of $S$ are independent. In particular, the lepton-flavour conserving entries determine the lepton-flavour violating ones, $S_{\ab} = \sqrt{S_{\alpha \alpha} S_{\beta \beta}}$. 
Note that these considerations exactly apply to $R_\ab$ as well, since the factors $\log(M_i/m_W)$ affect the overall scale in a flavour-universal way.

For $n_s>3$, there exists a factorised solution of \eq{Wto0} for $Y$, which is a straightforward generalisation of the cases $n_s=2,3$, with $n_s-2$ free complex parameters, $z_1,\dots,z_{n_s-2}$, in addition to the three real ones, $\lambda_{e,\mu,\tau}$.
However, such a factorised solution is no longer the most general, for we cannot in general write $Y$ in a compact form. 
For instance, the $n_s$ terms that contribute to \eq{Wto0} may cancel each other within separate subsets. 
In general, entries of $S$ obey a Cauchy-Schwarz inequality, $|S_\ab| \leq \sqrt{S_{\alpha \alpha} S_{\beta \beta}}$, since $S_\ab$ is a positive semi-definite matrix.
In particular, a diagonal entry $S_{\alpha \alpha}$ is zero if and only if $Y_{i\alpha} = 0$ for $i=1,\ldots , n_s$, and this implies $S_{\alpha\beta}=0$ for any $\beta$.
In contrast, the off-diagonal entries can vanish, suppressing LFV processes, while the diagonal ones are non-zero. These results hold for $R_\ab$ as well.

An example with $n_s = 4$ where LFV is suppressed is given by two Dirac pairs of sterile neutrinos with diagonal mass matrix $M = diag(M_1, M_1, M_2, M_2)$.
In the limit of unbroken lepton number $U(1)_L$, one finds
\begin{equation}
Y = \begin{pmatrix}
Y_{1\alpha} &
\pm i Y_{1\alpha} &
 Y_{2\alpha} &
\pm i Y_{2\alpha}
\end{pmatrix}^T
~~~
\Rightarrow~~~ 
S_{\ab} = \frac{2 Y_{1\alpha}^* Y_{1\beta}}{M_1^2} + \frac{2 Y_{2\alpha}^* Y_{2\beta}}{M_2^2} ~,
\label{2Dirac}\end{equation}
where only one of $Y_{1\alpha}$ and $Y_{2\alpha}$ can be taken to be real. 
Consequently, some off-diagonal entries of $S$ can vanish. 
For instance, imposing an additional $U(1)_{e }$ symmetry with $q_e(N_1)=1$ and $q_e(N_2)=0$, one finds $Y_{1\mu} = Y_{1\tau} = Y_{2e} = 0$, which implies $S_{e\mu} = S_{e \tau} = 0$, and all other entries non-vanishing. 
In this case, LFV occurs only in the $\mu-\tau$ sector.
Further imposing a full $U(1)_e \times U(1)_\mu \times U(1)_\tau$ symmetry, with e.g. $q_e(N_1)=1$, $q_\mu(N_2)=1$ and other charges vanishing,
only $S_{ee}$ and $S_{\mu\mu}$ are non-zero: there is no LFV and no corrections to the $\tau-\tau$ channel either.
Trivially, if one takes $n_s = 6$ with three Dirac pairs of sterile neutrinos,  one can make $S_\ab=0$ for any $\alpha\ne\beta$, while keeping $S_{\alpha\alpha}\ne 0$ for each $\alpha$.
In these examples, zeroes in $S$ correspond to zeroes in $R$ as well, because the cancellation is enforced by a symmetry.
More generally, it may happen that an accidental cancellation occurs in some $S_\ab$ but not in $R_\ab$, or vice versa, and in this case
LFV may appear only in log-enhanced WCs, or only in those with no logarithms.

We are now ready to discuss, in turn, each lepton observable that sets constraints on the seesaw parameters. 
The EFT framework elucidates the dependence of flavour-conserving (-violating) observables on the (off-) diagonal entries of the matrices $S$ or $R$.  
Therefore, the EFT approach enables a quick and direct comparison between bounds on different observables, in particular between those which conserve and violate flavour. 
We will begin in section \ref{EWobs} by discussing electroweak scale observables, that is, Higgs and $Z$ decays into leptons.
We will then discuss low-energy processes involving charged leptons: flavour-violating decays and scatterings in section \ref{LFV}, and  flavour-conserving observables
(dipole moments and tests of the universality of the Fermi constant) in section \ref{LFC}. 
We summarise all the constraints from flavour-conserving processes in table \ref{rtrace}, and all those from flavour-violating ones in table \ref{remu}, in terms of upper bounds
on the dimensionless parameters
\be
\hat{S}_\ab \equiv m_W^2 S_\ab~,\quad\quad \hat{R}_\ab \equiv m_W^2 R_\ab~.
\ee
In section \ref{plots} we will present summary plots for all the constraints, assuming the factorised solution for $\hat{S}_\ab$ and $\hat{R}_\ab$ where off-diagonal entries are determined by the diagonal ones.

\renewcommand{\arraystretch}{1.2}
\begin{table*}[t]
\centering
\begin{tabular}{|c|c|c|} \hline
Observable & Experimental value & Constraint \\ \hline \hline
$BR(Z \to \nu \nu)$ & $N_\nu = 2.9840 \pm 0.0082$ \cite{ALEPH:2005ab} & $1.15(\hat{S}_{ee} + \hat{S}_{\mu \mu}) + \hat{S}_{\tau \tau} \lesssim 3.5 \times 10^{-3}$ \\  \hline
$m_W$ & $80.379 \pm 0.012$ GeV \cite{Tanabashi:2018oca} & $\hat{S}_{ee} + \hat{S}_{\mu \mu} \lesssim 1.3 \times 10^{-3}$ \\  \hline
\hline
$\Gamma(Z \to e^+ e^-)$ & $83.92 \pm 0.12$ MeV \cite{ALEPH:2005ab}  &  $\hat{S}_{ee} + \hat{S}_{\mu \mu} \lesssim 0.53 \times 10^{-3}$  \\  \hline
$a_e^{\text{exp}} - a_e^{\text{SM}}$ 
&  $(-8.7 \pm 3.6) \times 10^{-13}$ \cite{Davoudiasl:2018fbb} & $\hat{S}_{ee} \lesssim 6.2$ \\ \hline
 \hline
$\Gamma(Z \to \mu^+ \mu^-)$ & $83.99 \pm 0.18$ MeV \cite{ALEPH:2005ab}  & $\hat{S}_{ee} + \hat{S}_{\mu \mu} \lesssim 1.4 \times 10^{-3}$ \\  \hline
$a_\mu^{\text{exp}} - a_\mu^{\text{SM}}$
& $(2.74 \pm 0.73) \times 10^{-9}$ \cite{Blum:2018mom} & $\hat{S}_{\mu \mu} \lesssim 0.13$ \\ \hline
\hline
$\Gamma(Z \to \tau^+ \tau^-)$ & $84.08 \pm 0.22$ MeV \cite{ALEPH:2005ab} & $\hat{S}_{ee} + \hat{S}_{\mu \mu} \lesssim 2.9 \times 10^{-3}$ \\  \hline
\hline
$G_F^{\mu \tau} / G_F^{e\tau}$ & $1.0018 \pm 0.0014$ \cite{Pich:2013lsa} & $\hat{S}_{ee}  \lesssim 2.6 \times 10^{-3}$ \\ \cline{1-2}
$G_F^{e\tau} / G_F$ & $1.0011 \pm 0.0015$ \cite{Pich:2013lsa} & $ \hat{S}_{\mu \mu} \lesssim 1.0 \times 10^{-3}$ \\  \cline{1-2} 
$G_F^{\mu\tau} / G_F$ & $1.0030 \pm 0.0015$ \cite{Pich:2013lsa} & $ \hat{S}_{\tau \tau} \lesssim 0.64 \times 10^{-3}$ \\ \cline{1-3} \hline
\end{tabular}
\caption{Experimental bounds on the seesaw parameters $\hat{S}_{\alpha \alpha}$ and $\hat{R}_{\alpha \alpha}$. For details on each bound see the relative section. }
\label{rtrace}
\end{table*}
%
%%%%%%%%%%%%%%%%%%%%%%%%%%%%%%%%%%%%%%%%%%%%%
%%%%%%%%%%%%%%%%%%%%%%%%%%%%%%%%%%%%%%%%%%%%%
%
\begin{table*}[h]
\renewcommand{\arraystretch}{1.2}
\centering
\begin{tabular}{|c|c|c|} \hline
Observable & Experimental upper limit & Constraint \\ \hline \hline
$BR(h \to e\mu)$ & $3.5 \lv{(0.3)} \times 10^{-4}$ $(95\%$ CL) \cite{Khachatryan:2016rke}\lv{ \cite{Calibbi:2017uvl}} & $|\hat{R}_{e\mu}| \lesssim 81 \lv{(24)}$ \\ \hline
$BR(Z \to e\mu)$ & $7.5 \times 10^{-7}$ $(95\%$ CL) \cite{Aad:2014bca} & $|\hat{R}_{e\mu}| \lesssim 0.065$ \\  \hline
$BR(\mu \to e\gamma)$ & $4.2 \lv{(0.6 )} \times 10^{-13}$ $(90\%$ CL) \cite{TheMEG:2016wtm}\lv{\cite{Baldini:2013ke}} & $|\hat{S}_{e\mu}| \lesssim 4.5 \lv{(1.7)} 
 \times 10^{-6}$ \\ \hline
$BR(\mu \to eee)$ & $1.0 \times 10^{-12} \lv{(10^{-16})}$ $(90\%$ CL) \cite{Bellgardt:1987du}\lv{\cite{Blondel:2013ia}} & $|\hat{R}_{e\mu}| \lesssim 5.6 \times 10^{-5} \lv{(5.6 \times 10^{-7})}$ \\ \hline
$BR(\mu  Au \to e  Au)$ & $7 \times 10^{-13}$ $(90\%$ CL) \cite{Bertl:2006up} 
& $|\hat{R}_{e\mu}| \lesssim 9.7 \times 10^{-6}$ \\ \hline
$BR(\mu  Ti \to e  Ti)$ & $4.3 \times 10^{-12} \lv{(10^{-18})}$  $(90\%$ CL) \cite{Dohmen:1993mp}\lv{\cite{Barlow:2011zza,Knoepfel:2013ouy}} 
& $|\hat{R}_{e\mu}| \lesssim 3.5 \times 10^{-5} \lv{(1.7 \times 10^{-8})}$ \\ \hline
\lv{$BR(\mu  Al \to e  Al)$} & \lv{$10^{-16}$ $(90\%$ CL) \cite{Kuno:2013mha}} 
& \lv{$|\hat{R}_{e\mu}| \lesssim 2.4 \times 10^{-7}$} \\ \hline
\hline
$BR(h \to e\tau)$ & $6.9 \lv{(0.3)} \times 10^{-3} $ $(95\%$ CL) \cite{Khachatryan:2016rke}\lv{ \cite{Calibbi:2017uvl}} & $|\hat{R}_{e\tau}| \lesssim 22 \lv{(4.5)}$ \\ \hline
$BR(Z \to e\tau)$ & $9.8 \times 10^{-6} $ $(95\%$ CL) \cite{Akers:1995gz} & $|\hat{R}_{e\tau}| \lesssim 0.24$ \\  \hline
$BR(\tau \to e\gamma)$ & $3.3 \lv{(0.5)} \times 10^{-8}$ $(90\%$ CL) \cite{Aubert:2009ag}\lv{\cite{Aushev:2010bq}} & $|\hat{S}_{e\tau} | \lesssim  3.0 \lv{(1.2)}  \times 10^{-3}$ \\ \hline
$BR(\tau \to eee)$ & $2.7 \lv{(0.05)} \times 10^{-8}$ $(90\%$ CL) \cite{Hayasaka:2010np}\lv{\cite{Aushev:2010bq}} & $|\hat{R}_{e\tau}| \lesssim 0.022$ \lv{$(3.0 \times 10^{-3})$} \\ \hline 
\hline
$BR(h \to \mu\tau)$ & $0.014 \lv{(3 \times 10^{-3})} $ $(95\%$ CL) \cite{Aad:2016blu}\lv{ \cite{Calibbi:2017uvl}} & $|\hat{R}_{\mu\tau}| \lesssim 31 \lv{(4.5)}$ \\ \hline
$BR(Z \to \mu\tau)$ & $1.2 \times 10^{-5} $ $(95\%$ CL) \cite{Abreu:1996mj} & $|\hat{R}_{\mu\tau}| \lesssim 0.26$ \\  \hline
$BR(\tau \to \mu\gamma)$ & $4.4 \lv{(0.3)} \times 10^{-8}$ $(90\%$ CL) \cite{Aubert:2009ag}\lv{\cite{Albrecht:2013wet}} & $|\hat{S}_{e\tau}| \lesssim  3.5 \lv{(0.9)}  \times 10^{-3}$ \\ \hline
$BR(\tau \to \mu \mu \mu)$ & $2.1 \lv{(0.1)} \times 10^{-8}$ $(90\%$ CL) \cite{Hayasaka:2010np}\lv{ \cite{Albrecht:2013wet}} & $|\hat{R}_{\mu \tau}| \lesssim 0.019$ \lv{($4.2 \times 10^{-3}$)} \\ \hline 
\hline
$|d_e|$ & $1.1 \times 10^{-29} e$ cm ($90\%$ CL) \cite{Andreev:2018ayy} & $| \text{Im}( \hat{S}_{e\mu} \hat{S}_{e\tau} \hat{S}_{\mu e}) | \lesssim 0.02$ \\ \hline
\end{tabular}
\caption{Experimental bounds on the seesaw parameters $\hat{S}_{\ab}$ and $\hat{R}_{\ab} $ for $\alpha \neq \beta$. Bounds in \lv{purple} are expected future limits.
For details on each bound see the relative section.}
\label{remu}
\end{table*}
Previous analyses of the seesaw phenomenology and the correlations among the various observables were performed e.g. in \cite{Antusch:2006vwa,Ibarra:2011xn,Alonso:2012ji,Fernandez-Martinez:2016lgt}, without using EFT techniques. 
An EFT analysis setting limits on the individual elements of $\hat{S}_\ab$ can be found in \cite{Abada:2007ux}.
The RG evolution of the theory was neglected, thus fewer observables were considered. 
Aspects of seesaw EFT phenomenology have also been studied in \cite{Broncano:2003fq,Cirigliano:2005ck,Antusch:2005gp,deGouvea:2007qla,Gavela:2009cd}.
We generalise and update these results by including lepton observables generated at leading-log, for arbitrary values of the seesaw parameters, and we highlight novel
correlations that sharply emerge from the EFT description. 
Recently improved experimental results for several observables also enable us to derive more stringent constraints on the seesaw parameter space than previous analyses.

%%%%%%%%%%%%%%%%%%%%%%%%%%%%%%%%%%%%%%%%%%%%%%%%%%%%%%
\subsection{Electroweak scale observables}
\label{EWobs}
%%%%%%%%%%%%%%%%%%%%%%%%%%%%%%%%%%%%%%%%%%%%%%%%%%%%%%

\subsubsection{Higgs boson decays into leptons}
\label{hdecay}
%%%%%%%%%%%%%%%%%%%%%%%%%%%%%%%%%%

We focus on Higgs boson decays into charged leptons since decays to neutrinos are suppressed by the smallness of the neutrino mass. 
Seesaw-induced corrections to flavour-conserving Higgs decays into quarks are also ignored as they lead to similar bounds, which are in any case overcome by more stringent bounds obtained in subsequent sections.

To derive the corrections to the charged lepton Yukawa couplings, one should account for the effects of several dim-6 operators induced by the seesaw at the electroweak scale $m_W$.
On one hand, after electroweak symmetry breaking the Higgs boson kinetic term receives corrections from $C^{HD}$ and $C^{H\square}$ (see e.g. \cite{Alonso:2013hga}), therefore one needs a field redefinition to restore a canonical kinetic term,\begin{equation}
h_{\text{SM}} = \left[1+\left(C^{H\square}-\frac14 C^{HD}\right) \frac{v^2}{\Lambda^2}  \right]h ~. %\simeq \left[ 1 + \frac{m_W^2 \text{tr}[R]}{8\pi^2} \log\left(\frac{M}{m_W}\right) \right] h ~,}
\label{hrescale}
\end{equation} 
On the other hand, the operator $Q^{eH}$ corrects both the charged lepton mass matrix and the charged lepton Yukawa couplings, 
\begin{align}
&{\cal L}_{\text{eff}}  \supset - \overline{e_{L\alpha}} \Bigg[\left(Y_{e,\ab} - C^{eH}_\ab \frac{v^2}{2\Lambda^2}\right) \frac{v}{\sqrt{2}} + \left(Y_{e,\ab} - C^{eH}_\ab \frac{3v^2}{2\Lambda^2}\right)  \frac{h_{\text{SM}}}{\sqrt{2}} \Bigg] e_{R\beta} + h.c. \nonumber \\
& = - \overline{e_{L\alpha}} \left[m_\alpha \delta_\ab  + \left(\frac{\sqrt{2}m_\alpha}{v}\delta_\ab - C^{eH}_\ab \frac{v^2}{\Lambda^2}\right) \frac{h_{\text{SM}}}{\sqrt{2}}\right]  e_{R\beta} + h.c. ,
\label{MvsY}
\end{align}
where in the third line we chose a basis where the charged lepton masses are diagonal.
Combining Eqs.~(\ref{hrescale}) and (\ref{MvsY}), one can extract the charged lepton Yukawa couplings to the physical Higgs boson, $h$, in the seesaw,
\begin{equation}
Y^h_{e,\alpha \beta} \simeq \left[\delta_\ab + \left( \frac{m_W^2 }{24\pi^2} \text{tr}[R] \delta_{\ab} - \frac{m_h^2}{16\pi^2 } R_{\alpha \beta} \right) \right] \frac{\sqrt{2}m_\beta}{v} ~,
\label{yuks}
\end{equation}
where we took the expressions for $C^{HD}$, $C^{H\square}$ and $C^{eH}$ in Eqs.~(\ref{HDmW}), (\ref{HsquaremW}) and (\ref{eHmW}), respectively, and expanded to retain only dim-6 corrections.

We can thus compute, at leading-log order, the Higgs decay widths. Let us begin with the LFV channels, $h \to \ell^+_\alpha \ell^-_\beta$ with $\alpha\ne \beta$.\footnote{
A detailed, model-independent analysis of these channels is provided in \cite{Blankenburg:2012ex,Harnik:2012pb}: in some cases they can be competitive with low-energy LFV processes.
We will show this is not the case in the seesaw.} 
Defining  $\Gamma(h \to \ell_\alpha \ell_\beta) \equiv \Gamma(h \to \ell_\alpha^+ \ell_\beta^-) + \Gamma(h \to \ell_\alpha^- \ell_\beta^+)$, 
one finds,
for $m_h\gg m_\beta \gg m_\alpha$, 
\begin{align}
\Gamma(h \to \ell_\alpha \ell_\beta) &\simeq \frac{m_h}{16\pi} \left| Y^h_{e,\alpha \beta} \right|^2 \simeq \frac{m^5_h m_\beta^2}{8\pi(16\pi^2)^2v^2} \left|R_{\alpha \beta}\right|^2  ~.
\label{widthhab}
\end{align}
Our EFT result agrees at leading-log order with an exact one-loop calculation in the inverse seesaw model \cite{Arganda:2017vdb}, as expected. 
In fact \cite{Arganda:2017vdb} the coefficient of the log-enhanced term is larger than the finite piece of the same order, which is $\mathcal{O}(Y^2/M^2)$, and for $|Y| \lesssim 0.3$ it is also larger than the $\mathcal{O}(Y^4/M^2)$ finite piece. 
Thus the leading-log estimate is accurate even for $M\sim$ TeV. 
The width of LFV Higgs decays in the seesaw was also computed in \cite{Pilaftsis:1992st,Korner:1992zk}. 
The LHC set upper bounds on the branching ratio of LFV Higgs decays  \cite{Tanabashi:2018oca}, assuming that the total Higgs production and width is SM-like.
In the seesaw, the latter receive corrections from dim-6 operators, which thus modifies the LFV branching ratios at dim-8 order and can thus be neglected.  
The corresponding bounds on $|\hat{R}_{\ab}|$ are feeble even for electroweak scale sterile neutrinos, see table \ref{remu}.

The Higgs decay width into same-flavour leptons is
\begin{align}
&\Gamma(h \to \ell_\alpha^+ \ell_\alpha^- ) \simeq \frac{m_h}{16\pi} \left| Y^h_{e,\alpha\alpha} \right|^2 \simeq \frac{m_h m_\alpha^2}{8\pi v^2} \Bigg| 1 + \frac{1}{16\pi^2} \left(\frac{2}{3} m_W^2 \text{tr}[R]  - m_h^2 R_{\alpha \alpha} \right) \Bigg|^2 .
\end{align}
Presently only $h \to \tau\tau$ has been observed \cite{Khachatryan:2016vau,Aaltonen:2013ioz}, with
branching ratio $[\sigma_h \cdot BR(h \to \tau \tau)^{\text{exp}}] = (1.12 \pm 0.23) [\sigma_h \cdot BR(h \to \tau \tau)^{\text{SM}}]$ \cite{Tanabashi:2018oca}. 
The seesaw modifies not only the $h \to \tau^+ \tau^-$ decay width, but also the Higgs production cross-section $\sigma_h$ 
and its total decay width $\Gamma_h$, both of which enter the experimental result. 
Even adding these corrections, the constraint on the linear combination of $\text{tr}[\hat{R}]$ and $\hat{R}_{\tau \tau}$ is too weak to be relevant.

\subsubsection{$Z$ boson decays into leptons}\label{Zdecays}
%%%%%%%%%%%%%%%%%%%%%%%%%%%%%%%%%%

We adopt the standard parametrisation for $Zf\overline{f}$ vector and axial couplings,
\begin{equation}
\mathcal{L}_Z = - \frac{e}{2s_wc_w} Z^\mu \overline{f_\alpha} \gamma_\mu \left( g^{V}_{f,\alpha \beta} - g^{A}_{f,\alpha \beta} \gamma_5 \right) f_\beta ~.
\label{Zcouplings}\end{equation}
The SM predicts $g^{V(\text{SM})}_{f,\alpha \beta} = [T_3(f_L)- 2s_w^2Q(f)] \delta_{\alpha \beta}$  and $g^{A(\text{SM})}_{f,\alpha \beta} = T_3(f_L) \delta_{\ab}$ at tree-level. 
The $Z$-boson couplings, $g^{V,A}$, are corrected by seesaw-induced dim-6 operators involving fermions. 
In contrast, the dim-6 operators that may directly correct gauge couplings and gauge boson kinetic terms are not induced by the seesaw at one-loop leading-log order.
Still, the $Z$-boson partial decay widths are indirectly affected by a tree-level shift of the Fermi constant, $G_F$. 
The latter is determined experimentally
from the decay $\mu \to e \overline{\nu_e} \nu_\mu$, with $G_F=1.166\times 10^{-5}$ GeV$^{-2}$. 
Due to the seesaw contribution to low-energy four-fermion operators (discussed in detail in section \ref{univL}), the measured quantity is  
\begin{equation}
\label{GFeq}
G_F \simeq G_F^{\text{SM}} - \frac{1}{4\sqrt{2}}  \left( S_{ee} + S_{\mu \mu} \right) ~ ,
\end{equation}
at linear order in $S_\ab$. 
The $Z$ partial widths are proportional to $G_F$ and, in addition, they depend on $s_w$, which can be expressed as a function of $\alpha$, $m_Z$ and $G_F$. While the experimental determination of the first two parameters is not significantly affected by the seesaw, 
the determination of $G_F$ is, therefore $s_w$ is shifted as well.
A useful, general discussion of the electroweak precision constraints on the SM EFT can be found e.g. in \cite{Falkowski:2014tna}.

Let us first consider $Z$-boson flavour-conserving decays to charged leptons. We find that the correction due to the shift in $G_F$ reads
\begin{align}
&\Gamma(Z \to \ell_\alpha^+ \ell_\alpha^-) \simeq \Gamma(Z \to \ell_\alpha^+ \ell_\alpha^-)^{\text{SM}}  \left[1 + \frac{v^2}{4} \frac{1 - 2 s_w^2 -4 s_w^4}{(1 - 2 s_w^2)(1-4s_w^2+8s_w^4)}({S}_{ee} + {S}_{\mu \mu}) \right] ,
\label{Zcharged}\end{align} 
where $\sin^22\theta_w \equiv (2\sqrt{2}\pi \alpha) /(G_F m_Z^2)$.\footnote{We have neglected the additional corrections to the $Z$-boson couplings to charged leptons, which arise via the seesaw RGEs,
because they are loop-suppressed: they will be relevant for $Z$-boson flavour-violating decays, see Eqs.~(\ref{gV})-(\ref{gA}).}
Comparing the SM predictions with the precise partial-width measurements made at LEP \cite{ALEPH:2005ab}, and allowing for 
a $2\sigma$ deviation, one reaches the stringent bound,
\begin{equation}
\hat{S}_{ee} + \hat{S}_{\mu \mu} \lesssim 0.53 \times 10^{-3} ~.
\end{equation}
As can be seen in table \ref{rtrace}, this is the strongest constraint on the flavour-conserving entries of $\hat{S}$. 
It comes from the measurement of $\Gamma(Z \to e^+ e^-)$, while the decays $Z \to \mu^+ \mu^-, \tau^+ \tau^-$ set comparable limits.

Let us discuss next the $Z$-boson invisible width.
The $Z$ couplings to neutrinos receive a correction from the WC $[C^{Hl(1)} - C^{Hl(3)}]$, which is induced at tree-level by the seesaw. 
The SM values, $g^{V(\text{SM})}_{\nu,\ab} = g^{A(\text{SM})}_{\nu,\ab} = \frac{1}{2} \delta_{\ab}$, are shifted by
\begin{equation}
\delta g^{V}_{\nu,\ab} = \delta g^{A}_{\nu,\ab} = - \frac{v^2}{2\Lambda^2} \left( C_{\alpha \beta}^{Hl(1)} - C_{\alpha \beta}^{Hl(3)}\right)  \simeq {-} \frac{v^2}{4} S_{\ab} ~.
\label{deltaznu}
\end{equation}
Combining this effect with the shift in $G_F$, we find that the effective number of light neutrinos is given by
\begin{align}
N_\nu  &\equiv \frac{\sum_\ab \Gamma(Z\to \nu_\alpha \overline{\nu_\beta})}{\Gamma(Z\to \ell^+_\gamma \ell^-_\gamma)}
\frac{\Gamma(Z\to \ell^+_\gamma \ell^-_\gamma)_{\text{SM}}}{\Gamma(Z \to \nu_\gamma \overline{\nu_\gamma})_{\text{SM}}} \nonumber\\
&\simeq 3 
- v^2 S_{\tau \tau} - \frac{1 - 3 s_w^2 + s_w^4 - 4 s_w^6}{(1 - 2s_w^2)(1 - 4s_w^2 + 8 s_w^4)} v^2 (S_{ee} + S_{\mu \mu})  ~,
%&\simeq 3 - v^2 {\rm tr}[S] - \frac{s_w^2 c_w^2(1 - 4 s_w^2)}{(1 - 2s_w^2)(1 - 4s_w^2 + 8 s_w^4)} v^2 (S_{ee} + S_{\mu \mu})  ~,
\label{znunu}
\end{align}
at linear order in $S_\ab$, and neglecting
one-loop suppressed seesaw corrections.
The coefficient of $(S_{ee}+S_{\mu\mu})$ differs from the one of $S_{\tau\tau}$ due to the shift in $G_F$. 
Thus our result is more accurate than in \cite{Akhmedov:2013hec}, where a flavour universal shift in $N_\nu$ is derived.
We also correct \cite{Fernandez-Martinez:2016lgt}, which uses a definition of $N_\nu$ different from the experimentally measured ratio of decay widths. 
The LEP measurement, $N_\nu = 2.9840 \pm 0.0082$ \cite{ALEPH:2005ab}, relies on measurements of the $Z$ total width and decay width into charged fermions. 
Demanding that $N_\nu$ is within the $2\sigma$ experimental interval sets a stringent bound, given in table \ref{rtrace}.

The seesaw also induces flavour-violating $Z$-boson couplings to charged leptons,
via the WCs $[C^{Hl(1)} + C^{Hl(3)}]$ and $C^{He}$, which arise at one-loop via the seesaw RGEs. 
We find, for $\alpha\ne\beta$,
\begin{align}
g^{V}_{\ell,\alpha \beta} &= {-} \frac{v^2 \left( C_{\alpha \beta}^{Hl(1)} + C_{\alpha \beta}^{Hl(3)} + C_{\alpha \beta}^{He} \right)}{2\Lambda^2} 
\simeq
 \frac{1}{16\pi^2} 
 %-\frac{{1-3t_w^2}}{3}  \text{tr}[\hat{R}] \delta_{\ab} {+} 
 \frac{17 +t_w^2}{6}  \hat{R}_{\alpha \beta}   ~,
\label{gV} \\
g^{A}_{\ell,\alpha \beta} &= {-} \frac{v^2\left( C_{\alpha \beta}^{Hl(1)} + C_{\alpha \beta}^{Hl(3)} - C_{\alpha \beta}^{He} \right) }{2\Lambda^2} \simeq
\frac{1}{16\pi^2}  
\frac{17 +t_w^2}{6}  \hat{R}_{\alpha \beta} ~.
\label{gA}
\end{align}
The width for LFV decays, defined by adding the $Z\to\ell_\alpha^+\ell_\beta^-$ and $Z\to\ell_\alpha^-\ell_\beta^+$ channels for $\alpha\ne\beta$, is given by 
\begin{align}
\Gamma(Z \to \ell_\alpha \ell_\beta) &\simeq \frac{m_Z^3}{6 \pi v^2} \left( |g_{\ell,\alpha \beta}^{V}|^2 + |g_{\ell,\alpha \beta}^{A}|^2  \right) \simeq \frac{ m_Z^3}{3 \pi v^2 (16\pi^2)^2} \left( \frac{17 +t_w^2}{6} \right)^2 \left| \hat{R}_{\alpha \beta} \right|^2 .
\label{widthZab}
\end{align}
Note that the shift of $G_F$, described by Eq. \eqref{GFeq}, affects the LFV width \eqref{widthZab} at higher order in $S$ only, thus it can be ignored.
Our result agrees at leading-log with a calculation in the inverse seesaw \cite{Herrero:2018luu}. 
The LFV $Z$-decay width in the seesaw was also computed in e.g. \cite{Illana:1999ww}. 
The experimental bounds \cite{Tanabashi:2018oca} translate into upper bounds on $\hat{R}_{\ab}$ which are summarised in table \ref{remu}. 

Eqs.~\eqref{widthhab} and \eqref{widthZab} imply, for $m_\beta \gg m_\alpha$,
\begin{equation}
\frac{BR(h \to \ell_\alpha \ell_\beta)}{BR(Z \to \ell_\alpha \ell_\beta)} \simeq \frac{3 m_h^5 m_\beta^2}{8m_Z^3m_W^4} \left( \frac{6}{17 +t_w^2} \right)^2 \frac{\Gamma_Z}{\Gamma_h}  \simeq 0.08\ \frac{m_\beta^2}{m_\tau^2} ~, 
%\simeq \frac{3 \lambda^2 y_\beta^2 v^2 m_h}{m_Z^3} \left( \frac{12}{g_1^2 + 17 g_2^2} \right)^2 = \simeq 1.3 y_\beta^2.
\label{hzratio}
\end{equation}
which is a sharp prediction of the seesaw at leading-log order. 
The two LFV decays are controlled by the same combination of seesaw parameters, but the Higgs boson decays are chirally-suppressed.
The present experimental sensitivity to LFV Higgs branching ratios is $\sim 10^3$ lower than for the $Z$, see table \ref{remu}.
We conclude that, in the seesaw, LFV Higgs decays are completely out of the experimental reach.

Finally, we note that flavour-conserving $Z$ decays to quarks are also shifted, due to the seesaw correction to $G_F$, while flavour-violating $Z$ decays are induced at one-loop, 
since the WCs $C^{Hq(1)}, C^{Hq(3)}, C^{Hu}, C^{Hd}$ are all generated at that order. 
However, decays to quarks are measured less precisely than leptonic ones, thus the limits are correspondingly weaker and we neglect them here.

%%%%%%%%%%%%%%%%%%%%%%%%%%%%%%%%%%%%%%%%%%%%%%%%%

\subsubsection{$W$ boson mass}

The seesaw correction to $G_F$ in \eq{GFeq}  also shifts the prediction of $m_W$, since the latter can be written as a function of $G_F$ and the other most precisely measured SM parameters, $\alpha$ and $m_Z$.\footnote{
We acknowledge Enrique Fernandez-Martinez for drawing our attention to this observable.
} 
One finds
\begin{equation}
m_W \simeq m_W^{\text{SM}} \left[1 + \frac{s_w^4 }{8 \pi \alpha (1- 2s_w^2) } (\hat{S}_{ee} + \hat{S}_{\mu \mu} ) \right] ~.
\label{mWshift}
\end{equation}
Here the SM prediction, including radiative corrections, is given by $m_W^{\text{SM}} = 80.362 \pm 0.008$ GeV \cite{deBlas:2016ojx}. 
The Eq.~\eqref{mWshift} is consistent with \cite{Loinaz:2004qc,Akhmedov:2013hec,Fernandez-Martinez:2016lgt}, and it should be compared with the very precise kinematic measurement of $m_W$ \cite{Tanabashi:2018oca}.
The corresponding bound on $(\hat{S}_{ee} + \hat{S}_{\mu \mu})$
is  reported in table \ref{rtrace}, where we allowed for a $2\sigma$ deviation between theory and experiment.

%%%%%%%%%%%%%%%%%%%%%%%%%%%%%%%%%%%%%%%%%%%%%%%%%%%%%%%%%%%%%%
\subsection{Low-scale flavour-violating observables}\label{LFV}
%%%%%%%%%%%%%%%%%%%%%%%%%%%%%%%%%%%%%%%%%%%%%%%%%%%%%%%%%%%%%%

\subsubsection{Charged lepton radiative decays}
%%%%%%%%%%%%%%%%%%%%%%%%%%%%%

As shown in sections \ref{matchM} and \ref{sec:matchmw}, the seesaw induces a non-zero electromagnetic dipole WC,  $C_{e\gamma, \alpha \beta}$, via one-loop matching at the scales $M$ and $m_W$, respectively.
The branching ratio of charged-lepton radiative decays is given by 
\begin{align}
BR(\ell_\alpha \to \ell_\beta \gamma) &\simeq 
\frac{m_\alpha^3 v^2}{8 \pi \Lambda^4 \Gamma_\alpha}
 \left( \left|C_{e\gamma,\alpha \beta} \right|^2 + \left|C_{e\gamma,\beta \alpha} \right|^2 \right) \simeq \frac{\alpha_{em} m_\alpha^5}{16  (16\pi^2)^2  \Gamma_\alpha} \left| S_\ab\right|^2 ~,
\label{raddec}
\end{align}
where $\Gamma_\alpha$ is the total width of $\ell_\alpha$, and in the second equality we replaced \eq{matchdipole} in the limit $C^W_\ab \to 0$.
Our result reproduces the original computation \cite{Cheng:1980tp,Altarelli:1977zq} at the lowest order in the matrix $\epsilon\equiv M^{-1}Yv/\sqrt{2}$, that is, in the limit where $m_\nu$ vanishes and the active-sterile mixing is approximated by $\epsilon$ 
(see appendix \ref{sec:ass} for a systematic derivation of higher orders,
corresponding to higher-dimensional operators in the EFT). 
The experimental bounds set very stringent constraints on $|\hat{S}_\ab|$ for $\alpha\ne\beta$, which we report in table \ref{remu}. 
The strongest one,  $BR(\mu \to e\gamma) < 4.2 \times 10^{-13}$ at $90\%$ C.L. \cite{TheMEG:2016wtm},
is expected to improve by an order of magnitude in the future \cite{Baldini:2013ke},
while radiative $\tau$ decays have branching ratios constrained to the $10^{-8}$ level.

It is interesting to study the correlation between charged-lepton radiative decays and LFV $Z$ decays discussed in section \ref{Zdecays}, which are log-enhanced. 
Indeed, Eqs.~\eqref{widthZab} and \eqref{raddec} imply
\begin{align}
\frac{BR(Z \to \ell_\alpha \ell_\beta)}{BR(\ell_\alpha \to \ell_\beta \gamma)} &
\simeq \frac{4m_Z^3m_W^4(17+t_w^2)^2}{27\pi v^2\alpha_{em} m_\alpha^5} \frac{\Gamma_\alpha}{\Gamma_Z} \frac{\left|\sum_i S_\ab^i \log \frac{M_i}{m_W}\right|^2}{|\sum_i S^i_\ab|^2} \simeq \frac{4m_Z^3m_W^4(17+t_w^2)^2 \Gamma_\alpha}{27\pi v^2\alpha_{em} m_\alpha^5\Gamma_Z} \ \log^2 \frac{M}{m_W}  \notag \\
&\simeq \log^2 \frac{M}{m_W} \times \begin{cases} 0.052~(\alpha=\tau) \\ 0.009 ~(\alpha=\mu) \end{cases} ,
\label{Zradratio}
\end{align}
where the second equality is accurate only in some limits, e.g. for $M_i\simeq M$ for all $i$, or for $|S^i_\ab|\gg|S^j_\ab|$ for all $j\ne i$ (in this case $M=M_i$).
The present experimental bounds imply that, in the $e-\mu$ ($e-\tau,~\mu-\tau$) channel, the constraint on $|\hat{R}_\ab|$ from $Z$ decays 
is about four (two) orders of magnitude weaker than the constraint on $|\hat{S}_\ab|$ from $\mu\to e\gamma$ ($\tau\to e\gamma,~\tau\to\mu\gamma$), see table \ref{remu}.
The only way to avoid this conclusion is to invoke a cancellation in $\sum_i S^i_\ab$, while different values of $\log(M_i/m_W)$ avoid the cancellation in the $Z$-decay
amplitude. 
As discussed below Eqs.~\eqref{rab2}-\eqref{2Dirac}, this is possible only for $n_s>3$.

\subsubsection{Lepton decays into three leptons}
%%%%%%%%%%%%%%%%%%%%%%%%%%%%%%%%%%%%

Another well-known flavour-violating process generated by the seesaw is $\ell_\alpha^- \to \ell_\beta^- \ell_\beta^+ \ell_\beta^-$ decays. 
The general expression for the branching ratio in EFT is given in \cite{Kuno:1999jp}. 
In terms of WCs generated by the seesaw after matching at $m_W$
(see section \ref{sec:matchmw}), we find, up to chirally-suppressed terms,
\begin{align}
BR(\ell_\alpha^- \to \ell_\beta^- \ell_\beta^+ \ell_\beta^-) &\simeq \frac{m_\alpha^5}{96\pi (16\pi^2) \Lambda^4 \Gamma_\alpha} \Bigg[ 8 \left| C^{V,LL}_{ee,\beta \ab \beta} \right|^2 + \left| C^{V,LR}_{ee,\beta \ab \beta} \right|^2 \notag\\
&+ \frac{32 e^2}{m_\alpha^2} \left( \log\frac{m_\alpha^2}{m_\beta^2} - \frac{11}{4} \right) \left| C_{e\gamma,\beta \alpha} \frac{v}{\sqrt2}\right|^2 
+ \frac{8e}{m_\alpha} \text{Re}\left( C_{e\gamma,\beta \alpha}^*\frac{v}{\sqrt{2}} \left( 4 C^{V,LL}_{ee,\beta \ab \beta}  + C^{V,LR}_{ee,\beta \ab \beta} \right)  \right) \Bigg]  
\notag \\
&\simeq \frac{m_\alpha^5 \left(27-96s_w^2+128s_w^4 \right)}{36 \pi v^4 (16\pi^2)^3 \Gamma_\alpha} \left| \hat{R}_{\alpha \beta} \right|^2 ~,
\label{3bodydec}
\end{align}
where in the final equality we neglected the contributions involving the dipole, as they are not log-enhanced (we checked that the log-enhanced terms dominate even for $M\sim$ TeV, and in any case we did not compute consistently the one-loop finite parts for the other WCs), and we used Eqs.~\eqref{matcheell} and \eqref{matcheelr}. 
This result agrees at leading-log with the highly non-trivial one-loop seesaw calculation of \cite{Ilakovac:1994kj} in the limit where $m_\nu$ vanishes. 
The corresponding bounds are collected in table \ref{remu}. The processes $\tau\to e(\mu^+\mu^-)$ and $\tau\to\mu(e^+e^-)$ give very similar bounds to $\tau\to3e$ and $\tau\to3\mu$, respectively.
Decays which violate flavour by two units,  $\tau^- \to \mu^+ e^- e^-$ and $\tau^- \to e^+ \mu^- \mu^-$, are not generated at leading-log order by dim-6 operators.

These rare decays are clearly correlated with other LFV decays, in particular
\begin{align}
\frac{BR(Z \to \ell_\alpha \ell_\beta)}{BR(\ell_\alpha^- \to \ell_\beta^- \ell_\beta^+ \ell_\beta^-)} &\simeq \frac{m_Z^3 v^2}{m_\alpha^5} \frac{16\pi^2(17+t_w^2)^2}{3(27-96s_w^2+128s_w^4)} \frac{\Gamma_\alpha}{\Gamma_Z} \simeq\begin{cases} 3.2~(\alpha=\tau) \\ 0.57 ~(\alpha=\mu) \end{cases}.
\end{align}
The experimental bounds on three-body decay branching ratios are much stronger (especially in the $e-\mu$ sector) than those from $Z$ decays, which are therefore completely out of reach as long as the leading-log approximation is accurate. 
Comparing with \eq{Zradratio}, one notices that $BR(\ell_\alpha\to 3\ell_\beta)$ can be as large as $BR(\ell_\alpha\to \ell_\beta\gamma)$ for $\log(M/m_W) \sim 8$.
The expected future limit $BR(\mu \to 3e) < 10^{-16}$ \cite{Blondel:2013ia}, four orders of magnitude tighter than the current bound \cite{Bellgardt:1987du}, should overcome the $\mu\to e\gamma$ constraint, see table \ref{remu}. The only more stringent bound may come from $\mu \to e$ conversion in nuclei, to which we turn now.

\subsubsection{The $\mu \to e$ conversion in nuclei}
\label{mutoeC}
The seesaw generates at one loop $2q2\ell$ operators, as well as the electromagnetic dipole operator, which both contribute to $\mu \to e$ conversion in nuclei. 
Recall we neglect $2q2\ell$ scalar operators, as they are Yukawa-suppressed, and retain only vector ones, see section \ref{sec:matchmw}. 
The $\mu \to e$ conversion rate, $\Gamma_N\equiv  \sigma(\mu N \to e N)$, is given by \cite{Kitano:2002mt,Cirigliano:2009bz} 
\begin{align}
\Gamma_N  &= \frac{m_\mu^5}{4 \Lambda^4 } \Bigg| \frac{D_N C_{e\gamma,e\mu}v}{\sqrt{2} m_\mu} +  4\sum \limits_{i=p,n}  V_N^i\sum \limits_{X=L,R} \Bigg( f_{Vi}^u C_{eu,e\mu uu}^{V,LX} + f_{Vi}^d C_{ed,e\mu dd}^{V,LX} \Bigg) \Bigg|^2 ,
\label{MueRate}
\end{align}
where the nucleon vector form factors are simply $f_{Vp}^u = 2, f_{Vp}^d = 1, f_{Vn}^u = 1, f_{Vn}^d = 2$, while the
nuclear form factors $D_N$ and $V_N^{p,n}$ are given in table \ref{properties}, for the nuclei that are most relevant for current or future bounds.
\begin{table}
\renewcommand{\arraystretch}{1.2}
\centering
\begin{tabular}{|c|c|c|c|} \hline
& ${}^{197}_{~79}$Au & ${}^{27}_{13}$Al & ${}^{48}_{22}$Ti \\ \hline
$D_N$ & 0.189 & 0.0362 & 0.0864 \\ \hline
$V_N^p$ & 0.0974 & 0.0161 & 0.0396 \\ \hline
$V_N^n$ & 0.146 & 0.0173 & 0.0468 \\ \hline
$\Gamma_N^{capt}$ [GeV] & $8.7 \times 10^{-18}$ & $4.6 \times 10^{-19}$ & $1.7 \times 10^{-18}$ \\ \hline
\end{tabular}
\caption{Nuclear form factors and capture rate for relevant nuclei \protect\cite{Kitano:2002mt}.}
\label{properties}
\end{table}
The matching in section \ref{sec:matchmw} gives
\begin{align}
\frac{C_{eu,e\mu uu}^{V,LL}}{\Lambda^2} + \frac{C_{eu,e\mu uu}^{V,LR}}{\Lambda^2} &\simeq \frac{1}{8\pi^2} \frac{64 s_w^2 - 27}{9 v^2} \hat{R}_{e\mu}  ~,
\label{muMatchu} \\
\frac{C_{ed,e\mu dd}^{V,LL}}{\Lambda^2} + \frac{C_{ed,e\mu dd}^{V,LR}}{\Lambda^2} &\simeq \frac{1}{8\pi^2} \frac{27 - 32 s_w^2}{9 v^2} \hat{R}_{e\mu} ~,
\label{muMatchd}
\end{align} 
while $C_{e\gamma,e\mu}$ is given in Eq.~\eqref{matchdipole} and is sub-leading as it is not log-enhanced.

Our result agrees at leading-log order with the explicit seesaw calculation performed in \cite{Alonso:2012ji}.
As pointed out in \cite{Alonso:2012ji}, the one-loop finite part is accidentally large and may cancel the log part for a tuned, nucleus-dependent value of $M$, typically around the TeV scale  (assuming all heavy neutrino masses are equal), e.g. $M = 4.7$ TeV for Titanium.
Only in this special case does the leading-log result give a poor estimate of the rate.

One can define a branching ratio, 
$BR(\mu N \to e N) \equiv \Gamma_N / \Gamma_N^{capt}$,
where $\Gamma_N^{capt}\equiv \sigma(\mu  N \to \nu_\mu N')$ is the muon capture rate. 
As shown in table \ref{remu}, the present experimental constraints on $\mu \to e$ conversion already give the strongest bound on $|\hat{R}_{e\mu}|$, which is very close to the bound on $|\hat{S}_{e\mu}|$ from $\mu \to e \gamma$. 
The limits from future $\mu \to e$ conversion experiments \cite{Kuno:2013mha,Barlow:2011zza,Knoepfel:2013ouy} are expected to be the most stringent ones. As usual, a cancellation in $|\hat{R}_{e\mu} |$ and not in $|\hat{S}_{e\mu}|$, or vice versa, 
cannot be excluded in the non-minimal scenarios with $n_s>3$, see the discussion below \eq{2Dirac}.

%%%%%%%%%%%%%%%%%%%%%%%%%%%%%%%%%%%%%%%%%%%%%%
\subsection{Low-scale flavour-conserving  observables}\label{LFC}
%%%%%%%%%%%%%%%%%%%%%%%%%%%%%%%%%%%%%%%%%%%%%%

\subsubsection{Magnetic dipole moments}
%%%%%%%%%%%%%%%%%%%%%%%%%%%%%%%%%%%%%%%%%%%%%%

The shift in the charged-lepton anomalous magnetic dipole moment, $a_\alpha$, is related to the electromagnetic dipole WC by
\begin{equation}
\Delta a_\alpha \equiv \frac{4m_\alpha v}{\sqrt{2} e\Lambda^2} \text{Re}(C_{e\gamma,\alpha \alpha})  \simeq - \frac{m_\alpha^2 S_{\alpha \alpha}}{16\pi^2} ~,
\label{MagMoment}
\end{equation}
where we replaced \eq{matchdipole} neglecting the loops of active neutrinos, which vanish as $m_\nu^2$.
Our EFT result agrees with the seesaw one-loop computation of \cite{Freitas:2014pua}. 
Besides the one-loop contribution given by \eq{MagMoment}, there may be higher-order corrections to $a_\alpha$ due to shifts induced by the seesaw on 
the Yukawa couplings, given in \eq{yuks}. In the SM computation of the magnetic moment, a Higgs boson exchange enters at one-loop order, inducing a shift $\Delta a_\alpha \sim \mathcal{O}[m_\alpha^2 R_{\beta\beta} /(16\pi^2)^2]$. This correction is loop-suppressed compared to Eq.~\eqref{MagMoment}. Moreover, as the consistency of the EFT requires $\hat{R}_{\beta\beta} \lesssim 1$, this correction is smaller than the current experimental precision on $\Delta a_\alpha$ for $\alpha=e,\mu,\tau$, thus we can safely ignore it.
%\bb{[here we discussed two-loop effects that are subleading; however we forgot one-loop effects due to the $G_F$ shift, that shifts the SM one-loop electroweak contribution to $a_\alpha$; as we explain in the following paper, these have similar size to \eq{MagMoment}; to include this here would require to extensively revise the text of this section; given that $a_\alpha$ is not a constraining observable for type-I seesaw, we ignore this story for this paper]}

The Eq.~\eqref{MagMoment} implies that the seesaw predicts a negative shift in the magnetic dipole moment of charged leptons, which is the opposite direction with respect to the $(g-2)_\mu$ anomaly, $a_\mu^{\text{exp}} - a_\mu^{\text{SM}} = (2.74 \pm 0.73) \times 10^{-9}$ 
\cite{Blum:2018mom} (see also \cite{Keshavarzi:2018mgv}).
A seesaw contribution of $\hat{S}_{\mu \mu} \simeq 0.07 k$ worsens the anomaly by $\sim k\sigma$: in table \ref{rtrace} we display the bound obtained taking $k=2$.
As the measurement of $Z \to \nu \overline{\nu}$ imposes the constraint $\hat{S}_{\mu \mu} \lesssim 3.5 \times 10^{-3}$, the seesaw correction to $(g-2)_\mu$ is negligible.

Recent improvements in the measurement of the fine-structure constant \cite{Parker191} and in the theoretical prediction for $(g-2)_e$ \cite{Aoyama:2017uqe} has led to a $2.4\sigma$ discrepancy in $(g-2)_e$, $a_e^{\text{exp}} - a^{\text{SM}} = (-8.7 \pm 3.6)  \times 10^{-13}$. 
This anomaly would be reduced by $1\sigma$ for $\hat{S}_{ee} \approx 1.4$ and it would fit for $\hat{S}_{ee} \approx 3.4$:
in table \ref{rtrace} we display the very weak $2\sigma$ upper bound on the seesaw contribution.
However these large corrections are ruled out by other constraints, most notably $Z \to \nu\overline{\nu}$. The size of the effect is rather suggestive of (non-seesaw) new physics close to or below the electroweak scale. 
Finally, the value of $a_\tau$ is poorly measured due to the very small $\tau$ lifetime, and it does not set any relevant constraint on $\hat{S}_{\tau \tau}$.

\subsubsection{Electric dipole moments}
\label{edms}
%%%%%%%%%%%%%%%%%%%%%%%%%%%%%%%%%%%%%%%%%%%%%%

The electric dipole moment (EDM) of charged leptons, $d_\alpha$, is related to the electromagnetic dipole WC by
\begin{equation}
d_\alpha \equiv - \frac{\sqrt{2}v}{\Lambda^2} \text{Im}\left(C_{e\gamma, \alpha \alpha}\right)~.
\end{equation}
In the seesaw, the one-loop contribution to $C_{e\gamma,\alpha\alpha}$, given by Eqs.~\eqref{EFTloopM}, \eqref{matchdipole} and \eqref{cegMatchMw}, is real, therefore the EDM vanishes at one loop.
Even beyond the dim-6 EFT approximation, the one-loop contribution remains real. 
We checked that two-loop diagrams contributing to the EDM must be finite.  
Indeed, applying the RGEs of \cite{Jenkins:2013zja,Jenkins:2013wua,Alonso:2013hga} twice to the seesaw WCs computed in section \ref{seesawEFT} does not induce terms of order $\sim (\alpha/4\pi)^2 \log^2(M/m_W)$ in $C_{e\gamma}$. 
The EFT contributions to the dipole of order  $\sim (\alpha/4\pi)^2 \log(M/m_W)$ are identified in \cite{Panico:2018hal}, and are not generated by the seesaw.\footnote{
This is consistent with the renormalisability of the seesaw Lagrangian: the lowest order contribution to the EDM must be finite, as there is no counter-term to cancel its presumed divergence.} 
Given the very stringent experimental constraint on $d_e$ \cite{Andreev:2018ayy}, finite two-loop contributions to the EDM may be phenomenologically relevant, and we estimate them below.

To find the leading contribution to the EDM, we must identify the shortest chain of Yukawa couplings that matches the transformation properties of the dipole bilinear, $(\overline{l_L}\sigma^{\mu\nu}e_R)$, and whose diagonal entries have a non-zero imaginary part. 
The anti-Hermitian part of such a chain is purely imaginary on the diagonal and thus gives the parametric form of the EDM. 
The minimal combination that satisfies these requirements is the commutator $[Y^\dagger Y Y_e^\dagger Y_e Y^\dagger Y, Y^\dagger Y]Y_e$ \cite{Ellis:2001yza,Smith:2017dtz}. 
For Dirac neutrinos ($M=0$), this is the whole story, however in the seesaw each pair $Y^\dagger Y$ is associated with a sterile neutrino exchange, which is integrated out at scale $M$,
therefore one must take the familiar replacement $Y^\dagger Y\to Y^\dag M^{*-1}M^{-1}Y=S$. It is an instructive exercise to check diagrammatically that of the 9 Higgs doublets associated with the 9 Yukawa couplings, at least 4 must be connected to form two loops. Thus, we obtain an estimate for the finite two-loop contribution to the electron EDM,
\begin{align}
|d_e| \sim \frac{2e}{(16 \pi^2)^2} \left( \frac{v}{\sqrt{2}} \right)^4 \text{Im}\left( \left[S Y_e^\dagger Y_e S, S \right]_{ee} \right) m_e  &= \frac{2em_ev^2}{(16 \pi^2)^2} (m_\tau^2-m_\mu^2) \text{Im}\left( S_{e\tau} S_{\tau\mu}S_{\mu e} \right) \notag \\
&\simeq 5.7 \times 10^{-28}\, \text{Im} \left( \hat{S}_{e \tau} \hat{S}_{\tau \mu} \hat{S}_{\mu e} \right) e \text{ cm}~.
\label{EDMe}
\end{align}
To our knowledge, this is the most accurate analytic estimate of the seesaw contribution to the electron EDM available in the literature.
It corresponds to a dim-10 operator in the seesaw EFT. 
The experimental upper bound on $d_e$ leads to the mild constraint $|\text{Im}(\hat{S}_{e\mu} \hat{S}_{e \tau} \hat{S}_{\mu \tau} )| \lesssim 0.02$, reported in table \ref{rtrace}.
The charged lepton EDMs were calculated in the seesaw in \cite{Abada:2015trh}, although it is difficult to compare our analytical estimate with their numerical results.

The stringent constraints on the flavour-violating parameters $|\hat{S}_\ab|$, imply a severe suppression of  this seesaw-induced EDM, $|d_e| \lesssim 10^{-37}e$ cm, comparable with the SM contribution
$|d^{\text{SM}}_e| \sim 10^{-38} e$ cm \cite{Pospelov:2013sca,Ghosh:2017uqq}. 
We note that for $n_s = 2,3$ sterile neutrinos, in the limit where light neutrino masses vanish, the matrix $S$ is real (see \eq{rab2} and \eq{rab3}) therefore the above contribution to the EDM vanishes.
Then a contribution to $d_\alpha$ not suppressed by $m_\nu$ can only be achieved via higher loops involving quarks, which are further Jarlskog suppressed, as in the SM. 
By contrast, for $n_s \geq 4$ the anti-Hermitian commutator $[S Y_e^\dagger Y_e S, S]_{\alpha \alpha}$ can be non-zero.

\subsubsection{Universality of lepton decays}
\label{univL}
%%%%%%%%%%%%%%%%%%%%%%%%%%%%%%%%%%%%%%%%%%%%%%%%%%%%%%%%%

The four-fermion Lagrangian which describes general $\ell_\delta \to \ell_\gamma \overline{\nu_\beta} \nu_\alpha$ decays is
\begin{align}
\mathcal{L} & \supset - \frac{4 G_F^{\text{SM}}}{\sqrt{2}} \left( \overline{\nu_\alpha} \gamma_\rho P_L \ell_\alpha \right) \left(\overline{\ell_\beta} \gamma^\rho P_L \nu_\beta \right)  + \frac{C^{V,LL}_{\nu e,\ab \gamma \delta}}{\Lambda^2} 
\left( \overline{\nu_\alpha} \gamma_\rho P_L \ell_\delta \right) \left( \overline{\ell_\gamma} \gamma^\rho P_L \nu_\beta \right),
\end{align}
with $C_{\nu e,\ab \gamma \delta}^{V,LL} = C_{\nu e,\beta \alpha \delta \gamma}^{V,LL*}$. This low-energy WC receives a tree-level contribution from the seesaw, given by Eq. \eqref{VLLnu}, which takes the form
\begin{equation}
\frac{C^{V,LL}_{\nu e,\ab \gamma \delta}}{\Lambda^2} \simeq \frac{1}{2} \left( S_{\alpha \delta} \delta_{\gamma\beta} + \delta_{\alpha \delta} S_{\gamma \beta} \right) ~.
\end{equation}
We will neglect seesaw one-loop corrections in the following. 
The neutrino flavour is not detected in experiments, thus the Fermi constant measured in $\mu \to e \nu \overline{\nu}$ decays is
\begin{align}
\label{GFeq1}
G_F^2 &\simeq \left| G_F^{\text{SM}} - \frac{1}{4\sqrt{2}}  \left( S_{ee} + S_{\mu \mu} \right) \right|^2 + \frac{1}{32} \left( 2| S_{e\mu} |^2 + |S_{e \tau}|^2 + | S_{\mu \tau}|^2 \right)  ~ .
\end{align}
This result was already displayed in \eq{GFeq} at linear order in $S_\ab$. 
Similarly, the effective Fermi constants for $\tau \to e \overline{\nu} \nu$ and $\tau \to \mu \overline{\nu} \nu$ decays are, respectively,
\begin{align}
(G_F^{e \tau})^2 &\simeq \left| G_F^{\text{SM}} - \frac{1}{4\sqrt{2}}  \left( S_{ee} + S_{\tau \tau} \right) \right|^2 + \frac{1}{32} \left( 2| S_{e\tau} |^2 + |S_{e \mu}|^2 + | S_{\mu \tau}|^2 \right) ~ , \\
(G_F^{\mu \tau})^2 &\simeq \left| G_F^{\text{SM}} - \frac{1}{4\sqrt{2}}  \left( S_{\mu \mu} + S_{\tau \tau} \right) \right|^2 + \frac{1}{32} \left( 2| S_{\mu\tau} |^2 + |S_{e \mu}|^2 + | S_{e \tau}|^2 \right) ~.
\end{align}

Bounds on the universality of $\ell_\alpha \to \ell_\beta \overline{\nu} \nu$ decays give \cite{Pich:2013lsa}
\begin{align}
&\frac{G_F^{\mu \tau}}{G_F^{e\tau}} -1 \simeq  \frac{ S_{e e} - S_{\mu \mu} }{4\sqrt{2} G_F} = 0.0018 \pm 0.0014
~,\label{univ1}\\
&\frac{G_F^{e\tau}}{G_F} -1 \simeq \frac{ S_{\mu \mu} - S_{\tau \tau} }{4\sqrt{2} G_F} = 0.0011 \pm 0.0015 
~,\label{univ2} \\
&\frac{G_F^{\mu\tau}}{G_F} - 1 \simeq \frac{ S_{e e} - S_{\tau \tau} }{4\sqrt{2} G_F} = 0.0030 \pm 0.0015 
~,\label{univ3} 
\end{align}
where we retained only terms linear in $S_\ab$. 
In table \ref{rtrace} we report the bounds on each $\hat{S}_{\alpha\alpha}$ from universality constraints, assuming the other entries are vanishing.
Since there is a $2\sigma$ discrepancy with the SM in \eq{univ3}, we conservatively use $3\sigma$ intervals from Eqs.~(\ref{univ1})-(\ref{univ3}) to set our bounds. 
It turns out that $G_F$ universality is a powerful constraint on $\hat{S}_{\alpha\alpha}$, comparable to or even slightly more stringent than the measurements of $m_W$ and $Z$-boson partial widths.
Note that the constraint is relaxed for $\hat{S}_{ee}\simeq \hat{S}_{\mu\mu} \simeq \hat{S}_{\tau\tau}$.

Lepton universality can be tested with comparable accuracy in pion and kaon leptonic decays \cite{Pich:2013lsa}. 
These bounds were exploited to constrain the seesaw parameters e.g. in \cite{Fernandez-Martinez:2016lgt}.
Here we restrict ourselves to purely leptonic observables, as the hadronic bounds are either weaker or of the same order.

%%%%%%%%%%%%%%%%%%%%%%%%%%%%%

%
\begin{figure}[t]
\begin{center}
\includegraphics[width=0.55\textwidth]{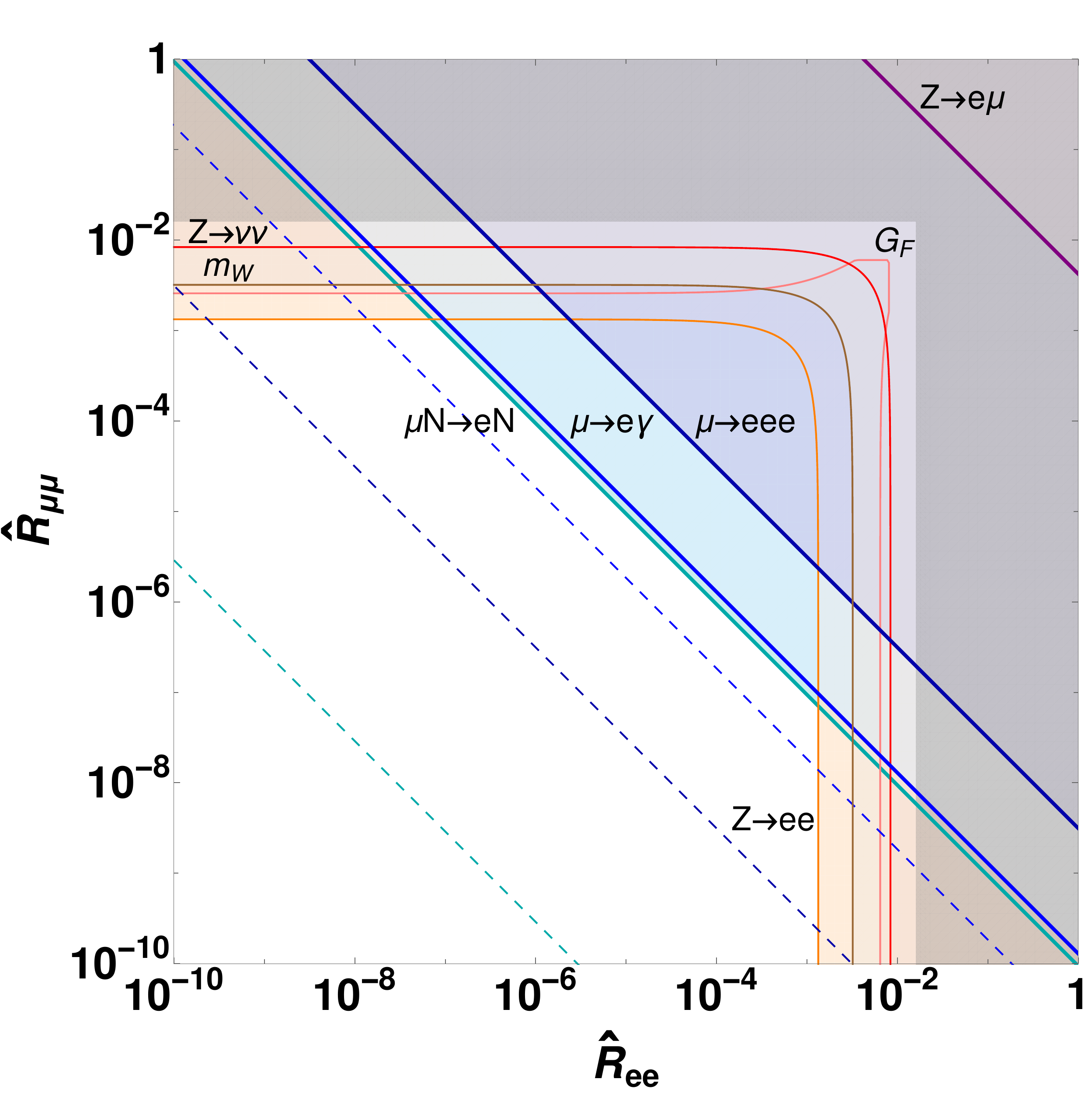}
\caption{\small Constraints on the seesaw EFT, in the $\hat{R}_{ee}$-$\hat{R}_{\mu \mu}$ plane, assuming $\hat{R}_{e\mu} = (\hat{R}_{ee} \hat{R}_{\mu \mu})^{1/2}$, $\hat{R}_{\alpha\tau}=0$ and $M_i = 1$ TeV.
The purple, dark blue, blue, and green solid lines represent bounds from LFV observables: $Z \to e \mu$, $\mu \to eee$, $\mu \to e \gamma$ and $\mu \to e$ conversion, respectively. 
The dashed lines of the same colours represent corresponding future sensitivities, where available. 
The orange, pink, brown and red lines represent bounds from flavour-conserving observables: $Z \to e^+ e^-$, $G_F$-universality, $m_W$, and $Z \to \nu \overline{\nu}$ respectively. 
As one enters the grey-shaded region, the validity of the EFT description becomes questionable.}
\label{emuPlot}
\end{center}
\end{figure}
%

%%%%%%%%%%%%%%%%%%%%%%%%%%%%%%%%%%%%%%%%%%%%%%%%%
%%%%%%%%%%%%%%%%%%%%%%%%%%%%%%%%%%%%%%%%%%%%%%%%%

\subsection{Summary plots} \label{plots}

%%%%%%%%%%%%%%%%%%%%%%%%%%%%%

In order to graphically compare the various constraints, we assume $S_\ab = \sqrt{S_{\alpha\alpha}S_{\beta\beta}}$ and $R_{\ab} = \sqrt{R_{\alpha \alpha} R_{\beta \beta}}$, which hold in general when $n_s=2$ or $3$.
Our results in the $e-\mu$, $e-\tau$ and $\mu-\tau$ sector are summarised in Figs.~\ref{emuPlot}, \ref{etauPlot} and \ref{mutauPlot}, respectively.
We plot the bounds as a function of $\hat{R}_{\alpha\alpha}$ and $\hat{R}_{\beta\beta}$, setting $\hat{R}_{\gamma\gamma}=0$, where $\alpha\ne\beta\ne\gamma$ are the three flavours. 
For $n_s>3$, the off-diagonal entries satisfy $|R_{\ab}| \le \sqrt{R_{\alpha \alpha} R_{\beta \beta}}$, see the discussion around \eq{2Dirac}. This means that the bounds from LFV observables shown in the figures can be relaxed to an arbitrary extent,relative to those from flavour-conserving observables.

In order to determine $\hat{S}_\ab$ as a function of $\hat{R}_\ab$, in the plots we fix the log factors by taking a unique seesaw scale, $M_i = 1$ TeV for all $i=1,\dots,n_s$.
This allows one to compare log-enhanced observables with those that do not carry a log.
As $M_i$ increases, the bounds with no log-enhancement become relatively weaker. 
Eq.~({\ref{Rdefn}) implies
\begin{align}
|\hat{R}_\ab|  &\simeq 0.016 \left|\sum_{i=1}^{n_s} Y_{i\alpha}^* Y_{i\beta}\right| \lesssim 0.016 \cdot n_s ~, 
\qquad ( M_i=1~{\rm TeV}~{\rm for}~i=1,\dots,n_s)~.
\end{align}
where the inequality is a conservative perturbativity bound, $|Y_{i\alpha}|\lesssim 1$.
Sterile neutrinos with mass below 1 TeV are also problematic in our approximation, as the one-loop leading-log corrections become comparable to the one-loop finite parts that we neglected.
Hence, in the figures the region $|\hat{R}_{\alpha \beta}|>0.016$ is shaded in grey, as the computability of our EFT becomes questionable.

\begin{figure}[p]
\begin{center}
\includegraphics[width=0.55\textwidth]{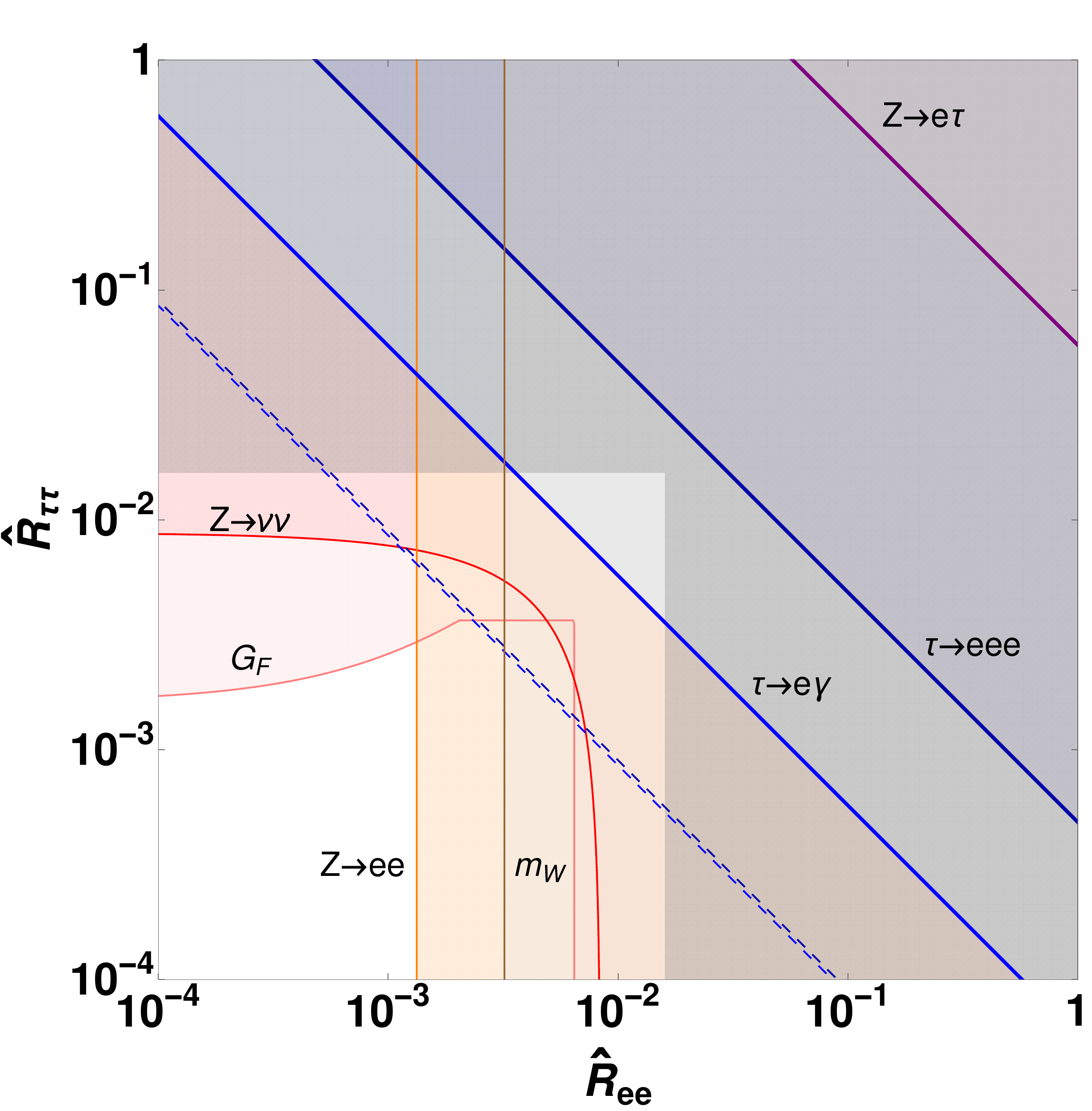}
\caption{\small Constraints on the seesaw EFT, in the $\hat{R}_{ee}$-$\hat{R}_{\tau \tau}$ plane, assuming $\hat{R}_{e\tau} = (\hat{R}_{ee} \hat{R}_{\tau \tau})^{1/2}$, $\hat{R}_{\alpha\mu}=0$ and $M_i = 1$ TeV.
The conventions are the same as in Fig.~\ref{emuPlot}.}
\label{etauPlot}
\end{center}
\end{figure}
\begin{figure}[p]
\begin{center}
\includegraphics[width=0.55\textwidth]{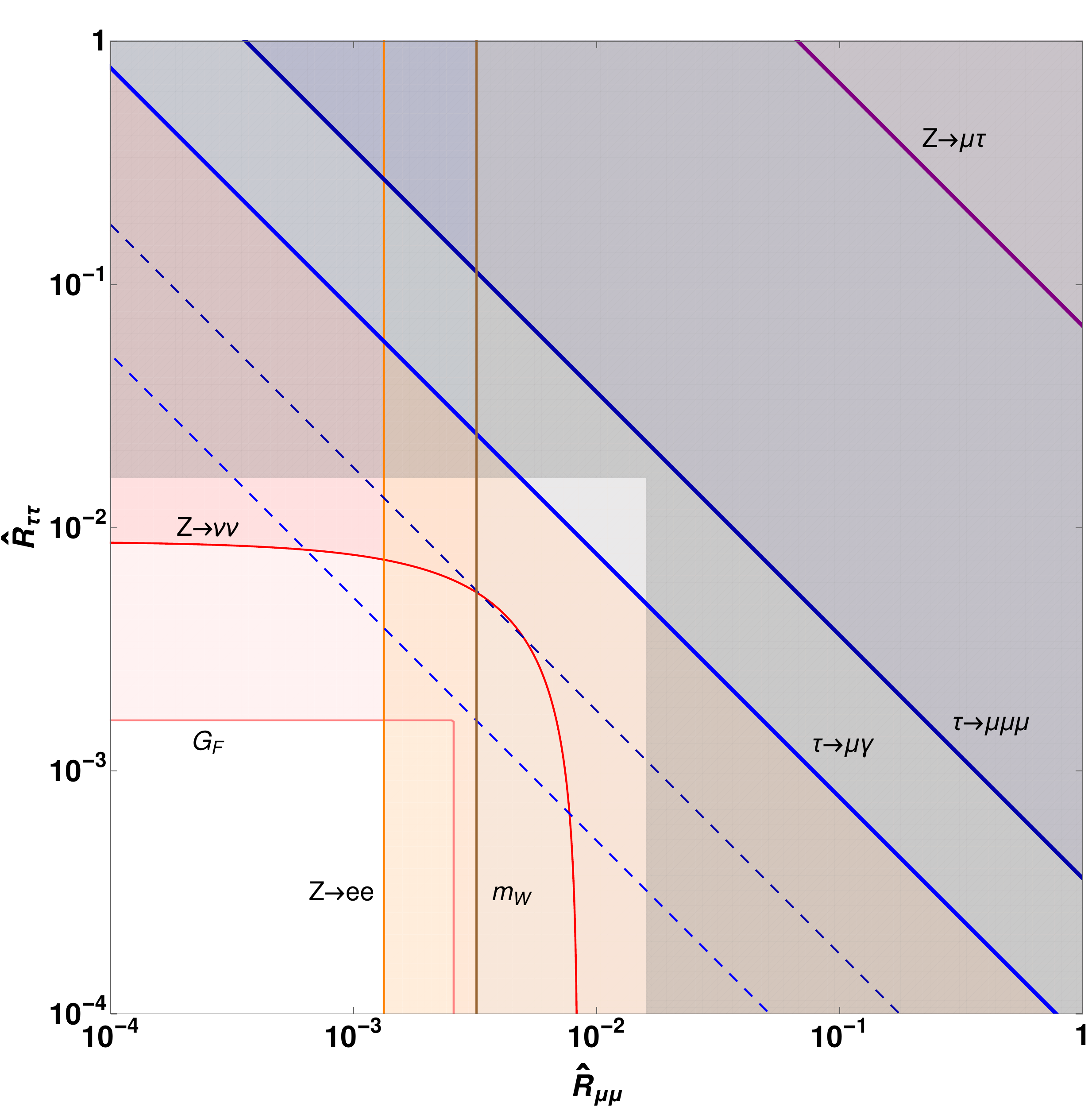}
\caption{\small Constraints on the seesaw EFT, in the $\hat{R}_{\mu\mu}$-$\hat{R}_{\tau \tau}$ plane, assuming $\hat{R}_{\mu\tau} = (\hat{R}_{\mu\mu} \hat{R}_{\tau \tau})^{1/2}$, $\hat{R}_{\alpha e}=0$ and $M_i = 1$ TeV.
The conventions are the same as in Fig.~\ref{emuPlot}.}
\label{mutauPlot}
\end{center}
\end{figure}

For the $e-\mu$ sector (Fig.~\ref{emuPlot}), the strongest bounds come from the various $\mu \to e$ transitions, as long as both $\hat{R}_{ee}, \hat{R}_{\mu \mu} \gtrsim 10^{-8}$.
Under the specified assumptions, the best limit comes from $\mu \to e$ conversion in gold and the strongest expected future bound is from $\mu \to e$ conversion in titanium. 
Constraints from flavour-conserving processes become dominant for either $\hat{R}_{ee}\lesssim 10^{-8}$ or $\hat{R}_{\mu \mu} \lesssim 10^{-8}$.
In this case the bound from $Z \to e^+ e^-$ is the tightest one.

For the $e-\tau$ (Fig.~\ref{etauPlot}) and $\mu - \tau$ (Fig.~\ref{mutauPlot}) sectors, we reach the striking conclusion that flavour-conserving bounds set the best limits on the seesaw parameters over the whole parameter range. 
This agrees qualitatively with the conclusions of \cite{Fernandez-Martinez:2016lgt}. 
Our figures show that even the future sensitivity of LFV $\tau$-decays is surpassed by present constraints from $G_F$-universality and $Z \to e^+ e^-$.
The dominance of flavour-conserving bounds is due to a combination of several factors: the flavour-conserving observables are induced at tree level, while LFV decays proceed at loop level; 
the seesaw Lagrangian implies that off-diagonal $R_\ab$ cannot be larger than $R_{\alpha\alpha}$; and the latter interferes with the SM couplings, while the off-diagonal $R_\ab$ do not. 
Indeed, only the extraordinary experimental precision of searches for $\mu \to e$ transitions (four to five orders of magnitude stronger than in corresponding searches of $\tau \to e,\mu$ transitions, see table \ref{remu}), 
enables flavour-violating bounds to become dominant in part of the $\hat R_{ee}-\hat R_{\mu\mu}$ plane.

We recall that our plots assume only two non-zero diagonal entries in $\hat{R}$. 
Suppose that the third is also non-zero. 
In this case, the bound from $Z \to \nu \overline{\nu}$ becomes stronger, since this observable constrains a weighted sum of the diagonal elements. 
The bounds from $m_W$ and $Z \to \ell_\alpha^+ \ell_\alpha^-$ in Figs. \ref{etauPlot} and \ref{mutauPlot} also become stronger with non-zero $R_{\mu\mu}$ and $R_{ee}$ respectively, as can be deduced from Table \ref{rtrace}.
The bound from $G_F$ becomes weaker, however, since it probes the difference of the diagonal entries.
In particular, if all three diagonal entries were equal, the $G_F$-universality constraint would disappear. 
On the contrary, constraints from flavour-violating observables are unaffected by the third diagonal entry.

Note that the bounds on $R$ and $S$ entries can be translated into a bound on the seesaw scales $M_i$, for any given choice of the matrix $Y$. 
If, for instance, one assumes that $Y$ is a matrix of order one numbers and that there is a unique seesaw scale, $M$, then $\mu \to e$ conversion in gold (titanium) places the strongest current (expected future) bound, $M \gtrsim 67$ $(2000)$ TeV.

%%%%%%%%%%%%%%%%%%%%%%%%%%%%%%%%%%%%%%
%%%%%%%%%%%%%%%%%%%%%%%%%%%%%%%%%%%%%%
%%%%%%%%%%%%%%%%%%%%%%%%%%%%%%%%%%%%%%
\section{Spurion analysis and perspective} 
\label{sec:spurions}
%%%%%%%%%%%%%%%%%%%%%%%%%%%%%%%%%%%%%%

\subsection{Implications of the seesaw flavour symmetry}\label{spu1}
%%%%%%%%%%%%%%%%%%%%%%%%%%%%%%%

We would like to investigate which predictions of the seesaw EFT can be derived by symmetry considerations only, and which depend on details of the matching and running procedure, or on numerical accidents.
To this end, it is enlightening to perform a spurion analysis.

As is well known, the lepton kinetic terms of the SM Lagrangian respect a large flavour symmetry, $G_L = SU(3)_{l_L} \times U(1)_{l_L} \times SU(3)_{e_R} \times U(1)_{e_R}$.
Once the charged-lepton Yukawa matrix $Y_e$ is introduced, $G_L$ is broken  to the product of lepton flavour numbers, 
$U(1)_{e} \times U(1)_{\mu} \times U(1)_{\tau}$.
The full symmetry is restored when $Y_e$ is treated as a spurion field, which transforms under $G_L$ as $Y_e \sim ( \overline{\mathbf{3}}_{-1} , \mathbf{3}_1)$, where
our notation is $({\bf R}_{Q}[l_L], {\bf R}_{Q}[e_R])$, with ${\bf R}$ the $SU(3)$ representations and $Q$ the $U(1)$ charges.
In the seesaw with $n_s$ sterile neutrinos, the lepton flavour symmetry is extended by an additional factor $SU(n_s)_{N_R} \times U(1)_{N_R}$. 
The spurion transformation of the neutrino Yukawa coupling is $Y \sim (\overline{\mathbf{3}}_{-1}, \mathbf{1}_0, \mathbf{F}_1)$, while the Majorana mass term transforms according to 
$M \sim (\mathbf{1}_0, \mathbf{1}_0, \mathbf{S}_2)$, where ${\bf F}$ (${\bf S}$) stands for the fundamental (two-index symmetric) representation of $SU(n_s)_{N_R}$, whose dimension is $n_s$ [$n_s(n_s+1)/2$].
The flavour symmetry assignments of fields and couplings are collected in table \ref{Transformations}.

The EFT operators $Q_i$ involving leptons transform non-trivially under $G_L$. 
By requiring that $C^i Q_i$ is invariant under $G_L$,
one can derive the spurion transformation of the WCs $C^i$ and, in turn, their parametric dependence on $Y_e$, $Y$ and $M$.
Some general rules apply.
Each coefficient $C^i/\Lambda^n$ must contain $n$ powers of $M^{-1}$ for dimensional reasons, where $n={\rm dim}(Q_i) - 4$.
As the entries of $Y_e$ are much smaller than one, only the lowest order in $Y_e$ is relevant.
If the entries of $Y$ are sufficiently smaller than one, a perturbative expansion in powers of $Y$ is also meaningful. 
Note that powers of $Y$ come necessarily in pairs, corresponding to the ``creation'' and ``annihilation'' of the sterile neutrino that is integrated out. 
In addition, since the EFT applies at low energy, i.e. for momenta $k\ll M$, every propagating sterile neutrino must cost at least one power of $M^{-1}$.
This means that the contraction $(Y^\dagger Y)_\ab$ is not allowed, and the sterile index in $Y_{i \alpha }$ must necessarily be contracted with a factor $M^{-1}_{ji}$.

In table \ref{Lepts} we display the representations of lepton bilinears,
and the associated operators. 
Let us start by applying the above prescriptions to lepton-number violating operators involving the bilinear $l_L l_L$.
It is easy to show that $C^W \sim Y^T M^{-1} Y$ is the unique dim-5 combination consistent with the seesaw flavour symmetry. 
This reproduces the EFT result of Eq.~\eqref{EFTtreeM}, up to the combinatorial factor $1/2$. 
At $\mathcal{O}(M^{-3} Y^2)$, there is one spurion that transforms as $\sim(\overline{\bf 6}_{-2},{\bf 1_0})$, namely $Y^T M^{-1} M^{-1*} M^{-1} Y$, however it is associated with dim-7 operators with two derivatives and so does not contribute to $m_\nu$.
At $\mathcal{O}(M^{-3}Y^4)$, there is
$C^{LH}/\Lambda^3\sim [ Y^T M^{-1} Y Y^\dagger M^{-1*} M^{-1} Y + (\dots)^T ]$,
which is associated with the dim-7 operator, $Q_{LH} \equiv Q_W(H^\dag H)$ \cite{Liao:2016hru,Lehman:2014jma}.
Note we have symmetrised in flavour space because the representation $\overline{\bf 6}$ is symmetric.
This induces a correction to the neutrino mass matrix,
$\Delta m_\nu \sim (v^4/4) C^{LH}/\Lambda^3$. 
The result fully agrees with the diagonalisation of the seesaw mass matrix at $\mathcal{O}(M^{-3})$, 
which we derived in appendix \ref{sec:ass}, up to an overall factor 1/2, see Eq.~\eqref{Eps4Appendix}. 

\begin{table}[]
\renewcommand{\arraystretch}{1.2}
\centering
\begin{tabular}{|c||c|c|c||c|c|c|}
\hline
& $l_L$ & $e_R$ & $N_R$ & $Y_e$ & $Y$ & $M$ \\ \hline \hline
$SU(3)_{l_L} \times U(1)_{l_L}$ & $\mathbf{3}_1$ & $\mathbf{1}_0$ & $\mathbf{1}_0$ & $\overline{\mathbf{3}}_{-1}$ & $\overline{\mathbf{3}}_{-1}$ & $\mathbf{1}_0$ \\ \hline
$SU(3)_{e_R} \times U(1)_{e_R}$ & $\mathbf{1}_0$ & $\mathbf{3}_1$ & $\mathbf{1}_0$ & $\mathbf{3}_1$ & $\mathbf{1}_0$ & $\mathbf{1}_0$ \\ \hline
$SU(n_s)_{N_R} \times U(1)_{N_R}$ & $\mathbf{1}_0$ & $\mathbf{1}_0$ & $\mathbf{F}_1$ & $\mathbf{1}_0$ & $\mathbf{F}_1$ & $\mathbf{S}_2$ \\ \hline
\end{tabular}
\caption{Transformation of lepton fields and couplings under the seesaw flavour symmetry.}
\label{Transformations}
\end{table}

Turning to lepton-number conserving operators, let us first consider the bilinear $\overline{l_L} l_L$, which transforms either as $(\mathbf{1}_0, \mathbf{1}_0)$ or $(\mathbf{8}_0, \mathbf{1}_0)$ under $G_L$, corresponding to the trace over flavour indices and the traceless part, respectively. The lowest order seesaw spurion with these transformation properties is 
$S_\ab = (Y^\dagger M^{-1*} M^{-1} Y)_\ab$, introduced in Eq.~\eqref{S}. The spurion $S$ has been extensively discussed in the seesaw literature, see e.g. 
\cite{Abada:2007ux,Gavela:2009cd,Smith:2017dtz}.
In section \ref{seesawEFT}, we showed that WCs associated with $\overline{l_L}l_L$ are indeed proportional to $S_\ab$ or ${\rm tr}(S)\delta_\ab$, or their log-enhanced versions, $R_\ab$ or ${\rm tr}(R)\delta_\ab$.
The bilinear $\overline{l_L} e_R$ transforms as $(\overline{\mathbf{3}}_{-1}, \mathbf{3}_1)$. The corresponding WCs receive a contribution 
$C^i/\Lambda^2 \sim S Y_e^\dagger$ or $R Y_e^\dag$, with a SM chiral suppression. 
The $\overline{e_R}e_R$ bilinear transforms as $(\mathbf{1}_0, \mathbf{1}_0)$ or $(\mathbf{1}_0, \mathbf{8}_0)$, with WCs which are doubly chiral-suppressed, $C^i/\Lambda^2 \sim Y_e S Y_e^\dagger$ or $Y_e R Y_e^\dag$.

\begin{table*}
\renewcommand{\arraystretch}{1.2}
\centering
\begin{tabular}{|c|c|c|}
\hline
Lepton bilinear & $G_L$ representation & SM EFT operators \\ \hline \hline
$\overline{l_L} l_L$ & ($\mathbf{1}_0 , \mathbf{1}_0$), $( \mathbf{8}_0 , \mathbf{1}_0 )$ & $Q_{ll}, Q_{Hl}^{(1,3)}, Q_{le}, Q_{lq}^{(1,3)},  Q_{lu}, Q_{ld}$ \\ \hline
$\overline{l_L} e_R$ & $(\overline{\mathbf{3}}_{-1} , \mathbf{3}_1)$ & $Q_{eH}, Q_{eB}, Q_{eW}, Q_{ledq}, Q_{lequ}^{(1,3)}$ \\ \hline
$\overline{e_R} e_R$ & $(\mathbf{1}_0 , \mathbf{1}_0)$, $( \mathbf{1}_0 , \mathbf{8}_0 )$ & $Q_{He}, Q_{ee}, Q_{le} , Q_{eu}, Q_{ed}, Q_{qe}$ \\ \hline
\hline
$l_L l_L$ & ($\mathbf{6}_2 , \mathbf{1}_0$), ($\overline{\mathbf{3}}_2 , \mathbf{1}_0$) & $Q_W$, $Q_{ll}$ \\ \hline
$l_L e_R$ & $(\mathbf{3}_1 , \mathbf{3}_1)$ & $Q_{le}$ \\ \hline
$e_R e_R$ & ($\mathbf{1}_0 , \mathbf{6}_2$), $( \mathbf{1}_0 , \overline{\mathbf{3}}_2 )$ & $Q_{ee}$ \\ \hline
\end{tabular}
\caption{Lepton bilinears and their transformation under the SM lepton flavour symmetry $G_L$. In the third column we list the dim-5 and dim-6 operators which contain each bilinear. 
}
\label{Lepts}
\end{table*}

At the next order in powers of $Y$, namely $Y^4 M^{-2}$, there is only one possible combination that transforms as $S_\ab$,
$4(C^{W\dag}C^W)_\ab/\Lambda^2 = (Y^\dagger M^{-1*} Y^* Y^T M^{-1} Y)_\ab$.\footnote{
Note we discount the spurion $Y^\dagger Y Y^\dagger M^{-1*} M^{-1} Y$,
which has the correct transformation properties, since a $Y^\dagger Y$ contraction is forbidden in the EFT below $M$, as already explained.}
This spurion, or rather its log-enhanced version, $\sum_\gamma W_{\ab\gamma\gamma}$, appears in various loop-suppressed WCs 
of section \ref{sec:RunningMmW}, sometime contracted with $Y_e$ for operators involving $e_R$.
Recall that, even for $Y\sim 1$, this spurion is necessarily very small as it is proportional to $m_\nu^2$.

Four-lepton operators transform as the product of two bilinears, for instance $Q_{ll} \sim [(\mathbf{1}_0, \mathbf{1}_0)  + (\mathbf{8}_0, \mathbf{1}_0)] \times \left[(\mathbf{1}_0, \mathbf{1}_0) + (\mathbf{8}_0, \mathbf{1}_0)\right]$.
At leading order, $C^{ll} \sim (R \delta + \delta R)$, which is reflected in Eq.~\eqref{llmW}. 
At the next order in powers of $Y$ and/or $M^{-1}$ there are pieces which transform under larger 
representations of $SU(3)_\ell$, specifically the $\mathbf{10}$, $\overline{\mathbf{10}}$, and $\mathbf{27}$. 
One example is provided by the $W_{\ab\gamma\delta}$ term in \eq{llmW}. 
These are negligible for our phenomenological purposes. 
A similar discussion applies for the other four-lepton operators, $Q_{le}$ and $Q_{ee}$. 

Finally, WCs of operators without leptons may be generated by the seesaw via a flavour-independent combination of spurions, $(\mathbf{1}_0, \mathbf{1}_0)$. At ${\cal O}(M^{-2})$, this invariant is obtained from $R_\ab$ or $W_{\ab\gamma\delta}$, 
by tracing over pairs of lepton indices, see e.g. Eqs.~(\ref{HDmW}) and (\ref{HsquaremW}).

Note that the leading-order dim-6 spurion $S_\ab$ is hermitian, so its diagonal entries are necessarily real. 
One can show that this is the case for dim-8 spurions as well, and complex flavour-diagonal WCs only appear at dim-10, and only for $n_s > 3$.
We have shown this explicitly for the dipole operator in section \ref{edms}.
Diagonal phases are present in lepton-number-violating WCs, such as $C^W_\ab$ or dim-7 WCs, but their overall size is generally suppressed by the smallness  of $m_\nu$.

We remark that the spurion analysis does not determine whether a given WC arises at tree level or at one loop, with or without log-enhancement, or at higher order. 
This is independent of the flavour symmetry, it depends on the gauge and Lorentz properties of the associated operator. 
For example, from symmetries one expects that $Z$ boson couplings to charged and neutral leptons are shifted at the same order. 
In reality, only the couplings to neutrinos are shifted at tree level, because the couplings to charged leptons accidentally cancel in the combination $[C^{Hl(1)} - C^{Hl(3)}]$.

%%%%%%%%%%%%%%%%%%%%%%%%%%%%%%%%%%%%%%%%%%%%%%%%%%%%%%%%%%%%%%%%%%%%%%%%%%%%%%%%%%%%%%%%%%%%%%%%%%
\subsection{Bottom-up analysis of lepton operators}\label{spu2}
%%%%%%%%%%%%%%%%%%%%%%%%%%%%%%

Let us now enlarge our analysis from the seesaw case to a generic new physics contribution to lepton operators. 
We will assume that the ultraviolet theory induces one (or more) spurion(s) in a definite
representation of the SM flavour symmetry, $G_L$, and derive the main phenomenological implications. 
These predictions will be common to any model that generates the given spurion. 
We do not aim for a general classification, but will rather choose some examples that have an intersection with the seesaw case, to allow a comparison with the results of the previous sections.

We begin by postulating the existence of a dimensionless spurion, $X \sim (\mathbf{3}_1,\mathbf{1}_0)$, amounting to a
coupling between a single SM lepton doublet and some new physics operator, $X\overline{l_L}{\cal O}_X$.
Notice that ${\cal O}_X$ can carry a lepton number (in the seesaw, $N_R$ can be assigned lepton number one), therefore the total lepton number $L_X$ of the spurion $X$ is arbitrary in general.
In particular, a WC for the Weinberg operator, $C^W_\ab \sim  X^*_{\alpha} X^*_{\beta}$, is allowed only for $L_X=1$,
and in this case, the size of the spurion is determined, $X^2 \sim m_\nu \Lambda /v^2$. 
In contrast, for $L_X\ne 1$, one needs an insertion of an additional $G_L$-singlet spurion, in order to match the lepton number of $C_W$ (e.g. in the seesaw, $M$ carries lepton number two).
In this case, the size of $X$ is not determined by $m_\nu$.
On the other hand, the WCs of lepton-number conserving operators are independent of $L_X$.
For the $(\mathbf{1}_0, \mathbf{1}_0)$ representation, one has
$C^i_\ab \sim X^*_{\gamma} X_{\gamma} \delta_\ab$, while for the $(\mathbf{8}_0, \mathbf{1}_0)$, one finds $C^i_\ab \sim (X^*_{\alpha} X_{\beta} - \frac{1}{3} X^*_{\gamma} X_{\gamma} \delta_\ab)$. 
The $(\overline{\mathbf{3}}_{-1}, \mathbf{3}_1)$ can also be induced, as $C^i_\ab\sim X^*_{\alpha}X_{\gamma}Y^*_{e,\beta\gamma}$. 
These dim-6 WCs can all lead to observable consequences for a sizeable $X$, i.e. $X \sim \Lambda/$TeV.
Their flavour structures are strongly correlated to each other.
For example, processes that require a chirality flip, such as $\ell_\alpha \to \ell_\beta \gamma$, are necessarily $Y_e$-suppressed, while those 
controlled by $(\mathbf{8}_0, \mathbf{1}_0)$, such as $\ell_\alpha \to 3\ell_\beta$, are not. Also, flavour-conserving and violating channels are strongly correlated, as
$|(X X^\dag)_\ab| = \sqrt{(X X^\dagger)_{\alpha \alpha} (X X^\dagger)_{\beta \beta}}$. 
Flavour violation by one unit, $\Delta F = 1$, arises at $\mathcal{O}(X^2)$, while $\Delta F = 2$ processes arise at $\mathcal{O}(X^4)$ and thus may be suppressed for small background values of the spurion. 
One may generalise these considerations to the case of more than one spurion in the same $G_L$ representation, $X_i \sim (\mathbf{3}_1,\mathbf{1}_0)$. Indeed, at least two are needed to induce realistic neutrino masses,
as the matrix $C^W_\ab$ should have rank two or larger.

Consider now a spurion with the quantum numbers of a lepton bilinear. 
The possibilities are listed in the second column of table \ref{Lepts}.
If one assumes that only the spurion $B_1 \sim (\mathbf{1}_0, \mathbf{1}_0)$ is present, no LFV is induced. 
Still, dim-6 WCs proportional to $B_1$ are constrained by flavour-conserving observables, especially $Z$ couplings to leptons and $G_F$-universality tests. 
The imaginary part of $B_1$ is strongly constrained by lepton EDMs, as the dipole operators are proportional to $B_1 Y_e^\dagger$.
As this spurion is a $G_L$-singlet, it can induce non-leptonic processes as well. 
Conversely, the spurion $B_8\sim (\mathbf{8}_0, \mathbf{1}_0)$ induces LFV, which strongly constrains its off-diagonal entries.  
Lepton flavour-conserving processes are subject to the condition of a traceless $B_8$, for instance  \eq{znunu} implies $N_\nu \geq 3$ when $\text{tr}[S] = 0$ (see section \ref{sec:pheno} for other phenomenological consequences). 
The vanishing trace also implies no corrections to non-leptonic operators. 
A spurion $B_{3\overline{3}} \sim (\mathbf{3}_{1}, \overline{\mathbf{3}}_{-1})$ directly generates operators containing the bilinear $\overline{l_L}e_R$, without any chiral-suppression. 
Dipole transitions strongly constrain the $B_{3\overline{3}}$ entries: the off-diagonal ones induce radiative charged-lepton decays and the diagonal ones correct magnetic and electric dipole moments.
 
Coming to lepton-number violating bilinears,  a spurion $B_6 \sim (\overline{\mathbf{6}}_{-2}, \mathbf{1}_0)$ may directly generate the Weinberg operator, provided its total lepton number is $L_{B_6} = -2$. 
In this case, its entries must be tiny to reproduce neutrino masses. 
If $L_{B_6}\ne -2$, one needs the insertion of an additional spurion to generate $C^W$, and $B_6$ entries may be large.
Then it becomes relevant to consider dim-6 WCs associated with two pairs of lepton doublets, $C^i \sim B^\dag_{6,\ab} B_{6,\gamma\delta}$, one pair, $C^i \sim (B_6^\dag B_6)_\ab$, and no pairs, $C^i \sim \text{tr}[B^\dag_6 B_6]$.
Finally, the spurion $B_3 \sim (\mathbf{3}_{-2},\mathbf{1}_0)$ is antisymmetric in its lepton doublet indices and therefore does not contribute to neutrino masses at leading order.
However, one can build $C^W \sim [B_3 Y_e^\dag Y_e + (...)^T]$, which may induce neutrino masses with a double chiral suppression.
The combination $B_3^\dag B_3$ can induce dim-6 WCs with a distinctive flavour structure.

Let us discuss how this bottom-up approach compares with the seesaw.
We showed that the two leading combinations of seesaw parameters that are singlets of $SU(n_s)_{N_R} \times U(1)_{N_R}$ are $C^W$ and $S$. 
Since $C^W \sim (\overline{\mathbf{6}}_{-2}, \mathbf{1}_0)$, it can be considered a spurion of type $B_6$ with total lepton number $-2$. 
It is indeed constrained by neutrino masses to be extremely tiny, therefore its effects on dim-6 operators, suppressed as $B_6^\dag B_6$, are negligible. 
The spurion $S$ transforms as a special combination of $(\mathbf{1}_{0}, \mathbf{1}_0)$ and $(\mathbf{8}_{0}, \mathbf{1}_0)$.
More precisely, recognising that $Y_{i\alpha}\sim X^*_{i,\alpha} \sim \overline{\mathbf{3}}_{-1}$ 
under $SU(3)_\ell \times U(1)_\ell$, where $X_i$ are $n_s$ spurions, one identifies the transformation properties
\be
C^W_\ab \sim X^*_{i,\alpha}X^*_{i,\beta}~,\quad\quad S_\ab \sim X_{i,\alpha}X^*_{i,\beta}~.
\ee
For $n_s > 1$ there can be cancellations among the $n_s$ contributions to $C^W$, possibly due to an approximate lepton number symmetry, while $S$ remains large.
Indeed, it is this observation which drives our phenomenological analysis in section \ref{sec:pheno}. 
An interesting inequality holds, $|(X_i X_i^\dag)_\ab| \le \sqrt{(X_i X_i^\dagger)_{\alpha \alpha} (X_i X_i^\dagger)_{\beta \beta}}$, 
that reproduces the inequality $|S_\ab| \leq \sqrt{S_{\alpha \alpha} S_{\beta \beta}}$ discussed at the start of section \ref{sec:pheno}. 
Finally, in the seesaw case the spurions that transform non-trivially under $SU(3)_{e_R}\times U(1)_{e_R}$ are necessarily proportional to one or more powers of $Y_e$.

We have shown that the seesaw model corresponds to a very specific set of spurions under the SM lepton flavour symmetry, $G_L$. Moreover, these spurions are not independent, 
rather they are specific combinations of the same set of Yukawa couplings and sterile neutrino masses.
The correlations are strictest for a small number of sterile neutrinos $n_s$. If a few deviations from the SM are discovered, besides neutrino oscillations, 
this pattern of correlations could be tested with some degree of confidence.

Alternative ultraviolet completions of the SM manifest themselves at low energy as different sets of spurions and correlations, therefore a qualitative comparison is possible without performing a detailed matching and running procedure. 
However, a precise comparison of two theories requires a computation of the full set of WCs, as we have done in this paper for the seesaw. 

A partial EFT treatment of alternative models of neutrino mass generation is available in the literature. For the type II and III seesaw, the tree-level EFT can be found in \cite{Abada:2007ux}. 
In the context of theories that address the gauge hierarchy problem, new physics close to the TeV scale may have a non-trivial interplay with neutrino mass generation and LFV. 
Such interplay has been studied with EFT and/or spurion techniques, for supersymmetric models e.g. in
\cite{Romanino:2001zf,Ellis:2001xt,Davidson:2006bd,Nikolidakis:2007fc,
Brignole:2010nh,Krauss:2011ur},
or in the compositeness scenario e.g. in \cite{Vecchi:2012fv,Redi:2013pga,Agashe:2015izu,Frigerio:2018uwx}. 
We believe it will be fruitful to apply our approach to these or other well-motivated models of new physics in the lepton sector, by performing a systematic comparison of the corresponding WCs.

\subsection{Summary of results} 

%%%%%%%%%%%%%%%%%%%%%%%%%%%%%%%%%%%%%%%%%%%%%%%%%

We developed the EFT of the seesaw in section \ref{seesawEFT} by implementing tree-level matching and one-loop running of dim-5 and dim-6 operators from the sterile neutrino mass scale, $M$, down to the energy scales of the observables.
The WCs are given in the leading-log approximation, but in appendix \ref{sec:RGEs} we display the one-loop RGEs, which may be used for a more accurate analysis of the running.
We also computed the WCs of the dipole operators by performing one-loop matching at the scales $M$ and $m_W$. This is essential to complete the EFT treatment of the seesaw.

This systematic EFT approach enabled us,  in section \ref{sec:pheno}, to consistently compute all relevant lepton observables at leading order. 
We started by demonstrating that the smallness of neutrino masses implies a very specific form of the neutrino Yukawa couplings, which in turn restricts the possible structures of the WCs:
there is an upper bound on the flavour-violating channels as a function of the flavour-conserving ones.
The bound is saturated for two or three sterile neutrinos, while flavour violation can be arbitrarily suppressed for $n_s>3$.

We identified which operators provide the leading contribution to each observable, and confronted the seesaw predictions with present and future experimental limits.
The EFT computation is arguably simpler than previous, direct one-loop computations.
The bounds are summarised in tables \ref{rtrace} and \ref{remu}, as well as in Figs.~\ref{emuPlot}, \ref{etauPlot} and \ref{mutauPlot}. 
The present experimental constraints are so tight that they completely exclude the grey-shaded region in those figures: this confirms the validity of our EFT approximations.

The EFT analysis highlights the correlations among the various observables.
On the LFV front, radiative and three-body decays of charged leptons give comparable constraints, and completely overcome searches for LFV in Higgs and even $Z$ decays. 
Limits on $\mu\to e$ conversion in nuclei are even tighter than LFV muon decays, especially in the long term.
Amusingly, LFV bounds also imply that the electron EDM must be extremely suppressed, as CP-violation is tied to flavour off-diagonal WCs.

Coming to flavour-conserving observables, besides the well-known bound from  $Z \to \nu \overline{\nu}$ (for which we fix some existing errors in the literature), 
we find even stronger constraints from $Z \to \ell_\alpha^+ \ell_\alpha^-$ decays, tests of $G_F$-universality in charged-lepton decays, and the precision measurement of $m_W$. 
The primacy of the $Z \to e^+ e^-$ bound on $S_{ee}$ and $S_{\mu \mu}$, as illustrated in Figs. \ref{emuPlot}, \ref{etauPlot} and \ref{mutauPlot}, has not been previously stated,  to our knowledge. 
These are the most stringent bounds on the seesaw parameters in the $\mu-\tau$ and $e-\tau$ sectors, where they overcome even future LFV searches. 
In the $e-\mu$ sector, the LFV probes are extremely sensitive, but the seesaw parameter space permits strong suppression of all WCs involving the electron with respect to the muon ones, or vice versa: in this case
$Z \to e^+ e^-$ becomes the ruling bound.

With a vast experimental programme expected to test lepton observables on many fronts, an understanding of the complementarity between them is very important to identify the ultraviolet theory from its low-energy footprints.
In sections \ref{spu1} and \ref{spu2} we investigated to what extent these footprints may allow one to distinguish the seesaw from a different model. 
We presented a detailed analysis of flavour symmetries, comparing the seesaw spurions with the most general ones, in order to underline the peculiarities of the seesaw EFT.
This illustrates the discriminating potential of our effective approach, and provides motivation to apply it to other models.

%%%%%%%%%%%%%%%%%%%%%%%%%%%%%%%%%%%%%%
\section*{Acknowledgements}
%%%%%%%%%%%%%%%%%%%%%%%%%%%%%%%%%%%%%%

We would like to thank S.~Davidson for several enlightening discussions.
This work is supported by the European Union's Horizon 2020 research and innovation programme 
under the Marie Sklodowska-Curie grant agreements No 674896 and No 690575, as well as 
by the OCEVU Labex (ANR-11-LABX-0060) and the A*MIDEX project (ANR-11-IDEX-0001-02) funded by the ``Investissements d'Avenir" French government program managed by the ANR.

%%%%%%%%%%%%%%%%%%%%%%%%%%%%%%%%%%%%%%
%%%%%%%%%%%%%%%%%%%%%%%%%%%%%%%%%%%%%%
\appendix
%%%%%%%%%%%%%%%%%%%%%%%%%%%%%%%%%%%%%%
%%%%%%%%%%%%%%%%%%%%%%%%%%%%%%%%%%%%%%

%%%%%%%%%%%%%%%%%%%%%%%%%%%%%%%%%%%%%%
\section{Seesaw diagonalisation} 
\label{sec:ass}
%%%%%%%%%%%%%%%%%%%%%%%%%%%%%%%%%%%%%%
\setcounter{equation}{0}
\renewcommand\theequation{A.\arabic{equation}}

In this appendix we will provide a systematic procedure for moving from the basis of active and sterile neutrinos, belonging to $SU(2)_L$ doublets and singlets respectively, to the basis
of light and heavy mass eigenstates. The diagonalisation of the seesaw matrix beyond the leading order has been already discussed using slightly different methods, e.g. in \cite{Grimus:2000vj} (which develops on \cite{Schechter:1981cv,Korner:1992zk}), and our results agree where they intersect.

Besides the general convenience of an accurate diagonalisation to study neutrino phenomenology, there are non-trivial connections with the EFT obtained by integrating out the sterile neutrinos, described in section \ref{seesawEFT}. 
It will be apparent that the tree-level WCs of operators with ${\rm dim}=4+n$ are related to the diagonalisation matrices at order $(M^{-1}m)^n$. 
Moreover, the diagonalisation is needed to compare the EFT prediction for a given observable, expressed in terms of operators involving only active neutrinos, and a computation of the same observable by Feynman diagrams that involves mass eigenstate neutrinos.

Let us begin by rewriting the mass terms in  \eq{lag} as
\be
{\cal L}_m = -\frac 12 \left(\overline{{\nu_L}^c}~\overline{N_R}\right)
\left(\ba{cc} 0 & m^T \\ m & M \ea\right) 
\begin{pmatrix} \nu_L \\ {N_R}^c \end{pmatrix} + h.c. ~,
\label{seesaw-diag}\ee
for an arbitrary number $n_a$ ($n_s$) of active (sterile) neutrinos.
We define a block diagonalisation of this symmetric mass matrix by
\begin{align}
{\cal M}&\equiv \left(\ba{cc} 0 & m^T \\ m & M \ea\right) = {\cal U}^* {\cal D} {\cal U}^\dagger  \equiv  \left(\ba{cc} V & W \\ X & Y \ea\right)^*      \left(\ba{cc} m_\nu & 0 \\ 0 & m_N \ea\right) \left(\ba{cc} V & W \\ X & Y \ea\right)^\dag ~,
\label{UD}
\end{align}
where the dimensions of the blocks are $(n_a\times n_a)$ for $V$ and $m_\nu$, $(n_s\times n_a)$ for $m$, $X$ and $W^T$, and $(n_s\times n_s)$ for $M$, $Y$ and $m_N$.
Here ${\cal U}$ is unitary, and the light and heavy mass matrices, $m_\nu$ and $m_N$, are diagonalised as
\be
m_\nu = U_\nu^* d_\nu U_\nu^\dag~, \qquad m_N = U_N^* d_N U_N^\dag~,
\ee
with $d_\nu$ and $d_N$ diagonal, real and positive, and $U_\nu$ and $U_N$ unitary. 
Working in the basis where the charged lepton masses are diagonal, the PMNS matrix, which describes the relation between active flavour eigenstates and light mass eigenstates, $\nu_{La} \equiv  (U_{\text{PMNS}})_{ai} \nu_{Li}$, takes the form
\be
U_{\text{PMNS}} = VU_\nu ~,
\ee
where $V$ is not unitary in general.
Note that the diagonalisation occurs in two steps: a unitary rotation ${\cal U}$, followed by a second one given by ${\rm diag}(U_\nu,U_N)$. This partition contains a degree of arbitrariness. It is natural to remove this ambiguity 
by requiring that ${\cal U}$ and ${\cal D}$ can be separately
expanded in powers of $m$ and $M$ only.
This convention guarantees e.g. that the light neutrino mass matrix at leading order is given by the canonical seesaw relation, $m_\nu=-m^TM^{-1}m$.

To ease the diagonalisation procedure, one can treat the various matrix blocks as spurions of the chiral symmetry
$U(n_a) \times U(n_s)$, which acts on the active and sterile neutrinos as $\nu_L\to U_a \nu_L$ and $N_R \to U_s N_R$ (do not confuse these symmetry transformations with the physical
unitary matrices involved in the diagonalisation). The corresponding spurion transformations are
\be
m\to U_s m U_a^\dag~,\qquad M\to U_s M U_s^T~.
\label{mMspur}\ee
In the convention where the matrices ${\cal U}$ and ${\cal D}$ of \eq{UD} can be separately
expanded in powers of $m$ and $M$, their blocks have to transform under $U(n_a) \times U(n_s)$ according to
\begin{align}
&V\to U_a V U_a^\dag~,& 
&W \to U_a W U_s^T~, &
& X\to U_s^* X U_a^\dag~,& \notag \\
&Y \to U_s^* Y U_s^T~,&
&m_\nu \to U_a^* m_\nu U_a^\dag~,&
&m_N \to U_s m_N U_s^T ~.&
\label{UDspur}
\end{align}
These relations restrict the possible combinations of $m$ and $M$ that can appear in the expansion of these blocks. 
Under the seesaw hypothesis, where the eigenvalues of $M$ are much larger than the entries of $m$, it is meaningful to determine the
matrices $V,W,X,Y$ as well as $m_\nu$, $m_N$ by an expansion in the dimensionless matrix (spurion)
\be
\epsilon\equiv M^{-1}m~,\qquad \epsilon \to U_s^* \epsilon U_a^\dag~.
\ee
It is then possible to solve \eq{seesaw-diag} order by order in $\epsilon$, by taking into account the unitarity condition ${\cal UU}^\dag = \mathbb{1}$ and by requiring the spurion transformations of \eq{UDspur} to hold.

For vanishing $\epsilon$ one has trivially 
\begin{align}
&V=\mathbb{1}~,&
&W=\mathbb{0}~,&
&X = \mathbb{0}~,& \notag \\
&Y=\mathbb{1}~,&
&m_\nu=\mathbb{0}~,&
&m_N =M~.&
\end{align}
At order $\epsilon$, the active-sterile mixing appears,
\be
W=\epsilon^\dag~,\qquad X=-\epsilon~.
\ee
At order $\epsilon^2$, non-unitary corrections to the PMNS matrix are generated, as well as the leading contribution to light neutrino masses, \begin{align}
&V=\mathbb{1}-\dfrac 12 \epsilon^\dag \epsilon~,&
&Y = \mathbb{1}-\dfrac12 \epsilon \epsilon^\dag ~,& \notag  \\
&m_\nu= -\epsilon^T M \epsilon~,&
&m_N =M +\dfrac12 \left(M\epsilon\epsilon^\dag+\epsilon^*\epsilon^TM\right)~.&
\end{align}
Note that even if one started with a basis where $M$ is diagonal, $m_N$ is no longer diagonal at this order, therefore $U_N$ is no longer the identity.
Note also the correspondence with the EFT of section \ref{seesawEFT}: $\epsilon^\dag \epsilon = Sv^2/2$ and $\epsilon^T M \epsilon = C^W v^2/\Lambda$.

At order $\epsilon^3$, one finds the next-to-leading correction to active-sterile mixing,
\be
W = -X^\dag = \epsilon^\dag -\dfrac 12 \epsilon^\dag\epsilon\epsilon^\dag -\epsilon^\dag M^* \epsilon^*\epsilon^T M^{-1*} ~.
\ee
At order $\epsilon^4$, it turns out that the separation between ${\cal U}$ and ${\cal D}$ is not uniquely defined. 
A unique solution is obtained by requiring that $V$ and $Y$ are hermitian, which is always possible by an appropriate choice of $U_\nu$ and $U_N$, respectively. 
In other words, the anti-hermitian correction to $V$ and $Y$ can be traded for a correction to $m_\nu$ and $m_N$, of the same order in $\epsilon$. 
We find
\be\def\arraystretch{2}\ba{l}
V=\mathbb{1}-\dfrac 12 \epsilon^\dag \epsilon +\dfrac 38 \epsilon^\dag \epsilon \epsilon^\dag\epsilon + \dfrac 12 \left[ \epsilon^\dag M^* \epsilon^* \epsilon^T M^{-1*} \epsilon + (...)^\dag \right]  ~,\qquad \\ 
Y = \mathbb{1}-\dfrac12 \epsilon \epsilon^\dag  + \dfrac 38 \epsilon \epsilon^\dag \epsilon \epsilon^\dag + \dfrac 12 \left[ \epsilon \epsilon^\dag M^* \epsilon^* \epsilon^T M^{-1*}  + (...)^\dag \right]   ~,\qquad  \\
m_\nu= -\epsilon^T M \epsilon +\dfrac 12 \left[\epsilon^T M \epsilon\epsilon^\dag\epsilon + (...)^T  \right]  ~,\qquad \\
m_N =M +\dfrac12 \Bigg[M\epsilon\epsilon^\dag -\dfrac 14 M\epsilon\epsilon^\dag\epsilon\epsilon^\dag - M \epsilon \epsilon^\dag M^* \epsilon^*\epsilon^T M^{-1*}  +(...)^T\Bigg] + \dfrac 14 \epsilon^*\epsilon^T M \epsilon\epsilon^\dag .
\label{Eps4Appendix}
\ea\ee
This provides, in particular, the next-to-leading contribution to light neutrino masses, corresponding to a dim-7 operator in the EFT.
If the ${\cal O}(\epsilon^2)$ contribution to $m_\nu$ vanishes, then the ${\cal O}(\epsilon^4)$ one vanishes as well.
In fact, it is remarkable that the condition $\epsilon^T M \epsilon \equiv m^T M^{-1} m = {\mathbb 0}$ is equivalent to the requirement ${\rm rank}({\cal M}) = n_s$,
and therefore it implies that $n_a$ neutrinos are massless at all orders. 
By contrast, if $\epsilon^\dag\epsilon$ vanishes, $V$ can still depart from the identity because $\epsilon\epsilon^\dag$ may be non-zero.  The ${\cal O}(\epsilon^4)$ corrections to $V$ correspond to dim-8 operators in the EFT.

Let us note that the dimensionless matrix $\epsilon$ may have entries not much smaller than one, e.g. for $M\sim 1$ TeV and $m\sim 100$ GeV, one has $\epsilon\sim 0.1$.
Therefore, next-to-leading corrections can be significant. 
They may also be the dominant effect if the leading contribution vanishes {\it and} the next one does not, as for the off-diagonal entries in $V$ and $Y$. 
We have derived above the next-to-leading correction for each block of the seesaw matrices, thus we refrain from displaying even higher orders in $\epsilon$. 

Finally, we remark that, in the spirit of the EFT, the seesaw diagonalisation should be performed at the largest mass scale, $\mu=M_{n_s}$, where ${\cal M}$ should be evaluated.
After the heaviest sterile neutrino has been integrated out, one should run down to $\mu=M_{n_s-1}$ and repeat the procedure, and so on and so forth. 
These threshold corrections to ${\cal M}$, due to the RG evolution from $M_{n_s}$ to $M_1$, are loop-suppressed and proportional to $\log(M_i/M_j)$.
While they are a sub-leading correction for the lowest order WCs, they may become significant compared to the higher powers of $\epsilon$ considered above.

%%%%%%%%%%%%%%%%%%%%%%%%%%%%%%%%%%%%%%
\section{RGEs for the seesaw effective operators} 
\label{sec:RGEs}
%%%%%%%%%%%%%%%%%%%%%%%%%%%%%%%%%%%%%%
\setcounter{equation}{0}
\renewcommand\theequation{B.\arabic{equation}}

\begin{table}[t]
\renewcommand{\arraystretch}{1.2}
\centering
\begin{tabular}{|c|c|} \hline
Name & Operator \\ \hline \hline
$Q_{W,\alpha \beta}$ & $(\overline{l_{L\alpha}^c}  \tilde{H}^* ) (\tilde{H}^\dagger l_{L\beta} )$ \\ \hline
\hline
$Q_{Hl,\alpha \beta}^{(1)}$ & $(\overline{l_{L\alpha}} \gamma_\mu l_{L\beta})(H^\dagger i \overleftrightarrow{D^\mu} H)$ \\ \hline
$Q_{Hl,\alpha \beta}^{(3)}$ & $(\overline{l_{L\alpha}} \gamma_\mu \sigma^A l_{L\beta})(H^\dagger i \overleftrightarrow{D^\mu} \sigma^A H)$ \\ \hline 
\hline
$Q_{eB,\alpha \beta}$ & $(\overline{l_{L\alpha}} \sigma_{\mu \nu} e_{R\beta}) H B^{\mu \nu}$ \\ \hline
$Q_{eW,\alpha \beta}$ & $(\overline{l_{L\alpha}} \sigma_{\mu \nu} e_{R\beta}) \sigma^A H W^{A \mu \nu}$ \\ \hline
\hline
$Q_{eH,\alpha \beta} $ & $(\overline{l_{L\alpha}} H e_{R\beta}) (H^\dagger H)$ \\ \hline
$Q_{He,\alpha \beta}$ & $(\overline{e_{R\alpha}} \gamma_\mu e_{R\beta})(H^\dagger i \overleftrightarrow{D^\mu} H)$ \\ \hline
$Q_{ll,\alpha \beta \gamma \delta}$ & $(\overline{l_{L\alpha}} \gamma_\mu l_{L\beta})(\overline{l_{L\gamma}} \gamma^\mu l_{L\delta})$ \\ \hline
$Q_{le,\alpha \beta \gamma \delta}$ & $(\overline{l_{L\alpha}} \gamma_\mu l_{L\beta})(\overline{e_{R\gamma}} \gamma^\mu e_{R\delta})$ \\ \hline
\hline
$Q_H$ & $(H^\dagger H)^3$ \\ \hline
$Q_{HD}$ & $(H^\dagger D_\mu H)^* (H^\dagger D^\mu H)$ \\ \hline
$Q_{H\square}$ & $(H^\dagger H) \square (H^\dagger H)$ \\ \hline \hline
$Q^{(1)}_{lq,\alpha \beta xy}$ & $(\overline{l_{L\alpha}} \gamma_\mu l_{L\beta})(\overline{q_{Lx}} \gamma^\mu q_{Ly})$ \\ \hline
$Q^{(3)}_{lq,\alpha \beta xy}$ & $(\overline{l_{L\alpha}} \gamma_\mu \sigma^A l_{L\beta})(\overline{q_{Lx}} \gamma^\mu \sigma^A q_{Ly})$ \\ \hline
$Q_{lu,\alpha \beta xy}$ & $ (\overline{l_{L\alpha}} \gamma_\mu l_{L\beta})(\overline{u_{Rx}} \gamma^\mu u_{Ry})$ \\ \hline
$Q_{ld,\alpha \beta xy}$ & $(\overline{l_{L\alpha}} \gamma_\mu l_{L\beta})(\overline{d_{Rx}} \gamma^\mu d_{Ry})$ \\ \hline
\hline
$Q_{uH,xy}$ & $(\overline{q_{Lx}} \tilde{H} u_{Ry}) (H^\dagger H)$ \\ \hline
$Q_{dH,xy}$ & $(\overline{q_{Lx}} H d_{Ry}) (H^\dagger H)$ \\ \hline
$Q^{(1)}_{Hq,xy}$ & $(\overline{q_{Lx}} \gamma_\mu q_{Ly})(H^\dagger i \overleftrightarrow{D^\mu} H)$ \\ \hline
$Q^{(3)}_{Hq,xy}$ & $(\overline{q_{Lx}} \gamma_\mu \sigma^A q_{Ly})(H^\dagger i \overleftrightarrow{D^\mu} \sigma^A H)$\\ \hline
$Q_{Hu,xy}$ & $(\overline{u_{Rx}} \gamma_\mu u_{Ry})(H^\dagger i \overleftrightarrow{D^\mu} H)$ \\ \hline
$Q_{Hd,xy}$ & $(\overline{d_{Rx}} \gamma_\mu d_{Ry})(H^\dagger i \overleftrightarrow{D^\mu} H)$ \\ \hline
\end{tabular}
\caption{\small List of the SM EFT operators induced by the seesaw, either at tree level or at one loop.}
\label{OpsSS}
\end{table}

Here we present the one-loop RGEs for the WCs of dim-5 and dim-6 operators which are induced by the type-I seesaw. 
The operators themselves are defined in table \ref{OpsSS}. 
The running of the Weinberg operator was derived in \cite{Chankowski:1993tx,Babu:1993qv,Antusch:2001ck}, the mixing of Weinberg squared into $d=6$ operators is taken from \cite{Davidson:2018zuo}, and we utilised \cite{Jenkins:2013zja,Jenkins:2013wua,Alonso:2013hga} 
for the mixing among $d=6$ operators. We adopt the conventions of the latter set of references, in particular  
the SM Yukawa couplings and Higgs potential are defined by 
\begin{align}
\mathcal{L}_{\text{SM}} &\supset - \overline{e_{R\alpha}} Y_{e,\alpha \beta} H^\dagger l_{L\beta} - \overline{d_{R\alpha}} Y_{d,\alpha \beta} H^\dagger q_{L\beta} - \overline{u_{R\alpha}} Y_{u,\alpha \beta} \tilde{H}^\dagger q_{L\beta} - \lambda \left(H^\dagger H - \frac{1}{2} v^2 \right)^2 ,
\label{SMLagrangian}
\end{align}
and the sign convention for the covariant derivatives is 
$D_\mu l_L \equiv [\partial_\mu + i g_1 (-1/2) B_\mu + i  g_2 (\sigma^a/2)W_\mu^a]l_L$, and similarly for the other fields.
The RGEs are calculated using dimensional regularisation in the $\overline{\text{MS}}$ scheme, as it is customary in EFT \cite{Georgi:1994qn}. 
We note that the one-loop anomalous dimensions are scheme-independent (as long as the chosen basis of operators is not redundant).
Scheme-dependence can arise at two-loop order (see e.g. \cite{Ciuchini:1993ks,Buras:1998raa}), which is beyond our scope.

In the type-I seesaw, the ultraviolet boundary conditions for the WCs are set by \eq{EFTtreeM}, that is, the only WCs different from zero are $C^W$ and $[C^{Hl(1)}- C^{Hl(3)}]$.
We neglect the RGE running induced by the dipole operators because it is a two-loop effect.
Then, the RGE for the Weinberg WC is given by
\begin{align}
16 \pi^2 \frac{dC_{ab}^W}{d\log\mu} &= - \frac{3}{2} (C^W Y_e^\dagger Y_e)_{ab} - \frac{3}{2} (Y_e^* Y_e^T C^W)_{ab} + 4 \lambda C^W_{ab} - 3 g_2^2 C^W_{ab} + 2 \chi C^W_{ab} ~,
\end{align}
where we defined
\begin{equation}
\chi \equiv \text{tr}\left[ 3 Y_u^\dagger Y_u + 3 Y_d^\dagger Y_d + Y_e^\dagger Y_e \right] \simeq 3 y_t^2 ~.
\end{equation}

The RGEs for the dim-6 operators involving leptons are 
\begin{align}
16 \pi^2 \frac{dC_{\alpha \beta}^{Hl(1)}}{d\log\mu} 
&= 2 (Y_e^\dagger Y_e C^{Hl(1)})_{\alpha \beta} + \frac{9}{2} (Y_e^\dagger Y_e C^{Hl(3)})_{\alpha \beta} + 2 \left( C^{Hl(1)} Y_e^\dagger Y_e \right)_{\alpha \beta} + \frac{9}{2} \left( C^{Hl(3)} Y_e^\dagger Y_e \right)_{\alpha \beta}  \notag \\
& + 2 \chi C_{ab}^{Hl(1)} + \frac{1}{3} g_1^2 C_{\alpha \beta}^{Hl(1)} + \frac{2}{3} g_1^2 \text{tr}[C^{Hl(1)}] \delta_{\alpha \beta} - 6 (C^{W\dagger} C^W)_{\alpha \beta} 
~,\\
16 \pi^2 \frac{dC_{\alpha \beta}^{Hl(3)}}{d\log\mu} 
&= \frac{3}{2} \left( Y_e^\dagger Y_e C^{Hl(1)} \right)_{\alpha \beta} + \left( Y_e^\dagger Y_e C^{Hl(3)} \right)_{\alpha \beta} + \frac{3}{2} \left( C^{Hl(1)} Y_e^\dagger Y_e \right)_{\alpha \beta} + \left( C^{Hl(3)} Y_e^\dagger Y_e \right)_{\alpha \beta} \notag \\
&+ 2 \chi C_{\alpha \beta}^{Hl(3)} + \frac{2}{3} g_2^2 \text{tr}[C^{Hl(3)}] \delta_{\alpha \beta} - \frac{17}{3} g_2^2 C_{\alpha \beta}^{Hl(3)} + 4 (C^{W\dagger} C^W)_{\alpha \beta} ~,\\
16 \pi^2 \frac{dC_{\alpha \beta}^{eH}}{d\log\mu} 
&= 4 \lambda \left( C^{Hl(1)} Y_e^\dagger + 3 C^{Hl(3)} Y_e^\dagger \right)_{\alpha \beta} + 2 \left(C^{Hl(1)} Y_e^\dagger Y_e Y_e^\dagger \right)_{\alpha \beta}\notag \\
&- 6 g_1^2 \left( C^{Hl(1)} Y_e^\dagger + C^{Hl(3)} Y_e^\dagger \right)_{\alpha \beta}
 - 4 \text{tr} \left[ C^{Hl(3)} Y_e^\dagger Y_e \right] (Y_e^\dagger)_{\alpha \beta} \notag \\
&+ \frac{4}{3} g_2^2  (Y_e^\dagger )_{\alpha \beta}  \text{tr}[C^{Hl(3)}] + 6 \left( C^{W\dagger} C^W Y_e \right)_{\alpha \beta} - 8 \text{tr} \left[ C^{W\dagger} C^W \right] ( Y_e^\dagger )_{\alpha \beta} ~,\\
16 \pi^2 \frac{dC_{\alpha \beta}^{He}}{d\log\mu} 
&= - 2 (Y_e C^{Hl(1)} Y_e^\dagger)_{\alpha \beta} + \frac{4}{3} g_1^2 \text{tr}[C^{Hl(1)}] \delta_{\alpha \beta} \\
16 \pi^2 \frac{dC_{\alpha \beta \gamma \delta}^{ll}}{d\log\mu} 
&=  \frac{1}{2} \left( C_{\ab}^{Hl(3)} - C_{\ab}^{Hl(1)} \right) (Y_e^\dagger Y_e)_{\cd} - C_{\alpha \delta}^{Hl(3)} (Y_e^\dagger Y_e)_{\gamma \beta} - \frac{1}{6} \left( g_2^2 C_{\ab}^{Hl(3)} + g_1^2 C_{\ab}^{Hl(1)} \right) \delta_{\cd} \notag \\
&+ \frac{1}{2} (Y_e^\dagger Y_e )_{\ab} \left( C_{\cd}^{Hl(3)} - C_{\cd}^{Hl(1)} \right)- (Y_e^\dagger Y_e)_{\alpha \delta} C_{\gamma \beta}^{Hl(3)}  - \frac{1}{6} \delta_{\ab} \left( g_1^2 C_{\cd}^{Hl(1)} + g_2^2 C_{\cd}^{Hl(3)} \right) \notag \\
& + \frac{g_2^2}{3} \left( C_{\alpha \delta}^{Hl(3)} \delta_{\gamma\beta } + C_{\gamma \beta}^{Hl(3)} \delta_{\alpha \delta} \right) - 2 C_{\alpha \gamma}^{W\dagger} C_{\beta \delta}^W ~,\\
16 \pi^2 \frac{dC_{\alpha \beta \gamma \delta}^{le}}{d\log\mu} 
&= 2 C_{\ab}^{Hl(1)} (Y_e Y_e^\dagger)_{\cd} - \frac{2g_1^2}{3} C_{\ab}^{Hl(1)} \delta_{\cd} ~.
\end{align}
This set of WCs controls Higgs and $Z$ boson decays to leptons, as well as charged lepton decays into three leptons, and corrections to $G_F$ universality. For $2q2\ell$ WCs, the RGEs read
\begin{align}
16 \pi^2 \frac{dC_{\alpha \beta \gamma \delta}^{lq(1)}}{d\log\mu} &=  C_{\alpha \beta}^{Hl(1)} (Y_u^\dagger Y_u - Y_d^\dagger Y_d )_{\gamma \delta} + \frac{g_1^2}{9} C_{\alpha \beta}^{Hl(1)} \delta_{\gamma \delta} ~,\\
16 \pi^2 \frac{dC_{\alpha \beta \gamma \delta}^{lq(3)}}{d\log\mu} &= - C_{\alpha \beta}^{Hl(3)} (Y_u^\dagger Y_u + Y_d^\dagger Y_d )_{\gamma \delta} + \frac{g_2^2}{3} C_{\alpha \beta}^{Hl(3)} \delta_{\gamma \delta} ~,\\
16 \pi^2 \frac{dC_{\alpha \beta \gamma \delta}^{lu}}{d\log\mu} &= - 2 C_{\alpha \beta}^{Hl(1)} (Y_u Y_u^\dagger)_{\gamma \delta} + \frac{4g_1^2}{9} C_{\alpha \beta}^{Hl(1)} \delta_{\gamma \delta} ~,\\
16 \pi^2 \frac{dC_{\alpha \beta \gamma \delta}^{ld}}{d\log\mu} &= 2 C_{\alpha \beta}^{Hl(1)} (Y_d Y_d^\dagger)_{\gamma \delta} - \frac{2g_1^2}{9} C_{\alpha \beta}^{Hl(1)} \delta_{\gamma \delta}  ~.
\end{align}
which are relevant to estimate $\mu\to e$ conversion on nuclei.
For operators with Higgs and quark fields, the RGEs are given by
\begin{align}
16 \pi^2 \frac{dC^{uH}_{xy}}{d\log\mu} &= - 4 \text{tr}[C^{Hl(3)} Y_e^\dagger Y_e] Y_{u,xy}^\dagger + \frac{4}{3}g_2^2 \text{tr}[C^{Hl(3)}] Y_{u,xy}^\dagger ~, \\
16 \pi^2 \frac{dC^{dH}_{xy}}{d\log\mu} &= -4 \text{tr}[C^{Hl(3)} Y_e^\dagger Y_e] Y_{d,xy}^\dagger + \frac{4}{3}g_2^2 \text{tr}[C^{Hl(3)}] Y_{d,xy}^\dagger ~,\\
16 \pi^2 \frac{dC^{Hq(1)}_{xy}}{d\log\mu} &= -\frac{2}{9} g_1^2 \text{tr}[C^{Hl(1)}] \delta_{xy} ~,\\
16 \pi^2 \frac{dC^{Hq(3)}_{xy}}{d\log\mu} &= \frac{2}{3} g_2^2 \text{tr}[C^{Hl(3)}] \delta_{xy} ~,\\
16 \pi^2 \frac{dC^{Hu}_{xy}}{d\log\mu} &= - \frac{8}{9} g_1^2 \text{tr}[C^{Hl(1)}] \delta_{xy} ~,\\
16 \pi^2 \frac{dC^{Hd}_{xy}}{d\log\mu} &= \frac{4}{9} g_1^2 \text{tr}[C^{Hl(1)}] \delta_{xy} ~.
\end{align}
These WCs induce a small shift in the Higgs and $Z$ boson couplings to quarks, which we neglected as they are typically less constraining than their lepton counterparts.

Finally, the RGEs for operators with Higgs fields and derivatives only are
\begin{align}
16 \pi^2 \frac{dC^H}{d\log\mu} &= \frac{16}{3} \lambda g_2^2 \text{tr}[C^{Hl(3)}] 
- 16 \lambda \text{tr}[C^{Hl(3)} Y_e^\dagger Y_e] - 32 \lambda \text{tr} [ C^{W\dagger} C^W] \label{CH} ~,\\
16 \pi^2 \frac{dC^{HD}}{d\log\mu} &= - 8 \text{tr}[C^{Hl(1)} Y_e^\dagger Y_e] - \frac{8}{3} g_1^2 \text{tr} [ C^{Hl(1)}] - 16 \text{tr}[C^{W\dagger} C^W] ~,\\
16 \pi^2 \frac{dC^{H\square}}{d\log\mu} &= - 2 \text{tr}[(3C^{Hl(3)} + C^{Hl(1)}) Y_e^\dagger Y_e] + 2 g_2^2 \text{tr}[ C^{Hl(3)}] - \frac{2}{3} g_1^2 \text{tr}[C^{Hl(1)}] - 8 \text{tr}[C^{W\dagger} C^W] ~.
\end{align}
Our result for the last term of \eq{CH} is a factor of two smaller than the corresponding term in \cite{Davidson:2018zuo}.

Note that dim-6 operators may also mix into dim-4 operators, as the SM contains a dim-2 operator, $H^\dagger H$. 
Consequently, the non-zero $[C_{Hl}^{(1)}-C_{Hl}^{(3)}]$ generated at tree-level by the seesaw introduces corrections to the $\beta$-functions of the SM parameters $\lambda$ and $Y_e$. 
These effects are sub-leading since they are suppressed by $v^2/M^2 \ll 1$ \cite{Jenkins:2013zja}.

The set of RGEs presented in this appendix, together with the RGEs for the SM couplings (see e.g. \cite{Buttazzo:2013uya}), are of course coupled to each other.
Therefore, during the evolution from $M$ to $m_W$, the running of each WC is affected, at next-to-leading order, by the running of the SM couplings and the other WCs.
Let us roughly estimate the size of these corrections in the seesaw.
The largest and fastest-running couplings in the SM are $y_t$ and $g_3$, which do not enter into the most relevant WCs, see Eqs.~\eqref{HlmW}\textendash \eqref{HsquaremW}
(the $2q2\ell$ WCs pertinent for $\mu \to e$ conversion do not involve the top quark). 
The  most relevant running is that of the Higgs quartic coupling, with $\beta_\lambda = d\lambda/d\log\mu \simeq -3 y_t^4/(8\pi^2)$. 
The running of $g_1,g_2$ is much weaker, $\beta_{g} \sim g^3/(16\pi^2)$.
The seesaw tree-level WC, $C^{Hl} \equiv [C^{Hl(1)}-C^{Hl(3)}]/2$, has also a strong scale-dependence, since $\gamma_{Hl} \equiv [
\gamma^{Hl(1)}_{Hl(1)} + \gamma^{Hl(3)}_{Hl(3)}- \gamma^{Hl(1)}_{Hl(3)} - \gamma^{Hl(3)}_{Hl(1)}]/2 \simeq 3y_t^2 /(8\pi^2)$,
where the anomalous dimensions are defined by \eq{RGEs}.

To estimate the dominant correction to the value of the WCs at $m_W$, let us
consider the running of $C^i$ due to $C^{Hl}$, 
with the assumption that $\gamma^i_{Hl}$ does not depend on rapidly-running SM couplings.
Then, the solution of \eq{RGEs}  reads
\begin{align}
C^i(m_W) & \simeq C^i(M) - \int_{\log m_W}^{\log M} {\rm d}\log\mu \ \gamma^i_{Hl} C^{Hl} (\mu) \simeq C^i(M) - \int_{C^{Hl}(m_W)}^{C^{Hl}(M)} {\rm d}C^{Hl} \frac{\gamma^i_{Hl}}{\gamma_{Hl}} ~,
\end{align}
where $C^{Hl}(m_W) \simeq C^{Hl}(M) (m_W/M)^{\gamma_{Hl}}$. 
A perturbative expansion gives
\begin{align}
C^i(m_W) \simeq C^i(M) - \gamma^i_{Hl} C^{Hl}(M) \log \frac{M}{m_W} \left(1+ \frac{\gamma_{Hl}}{2} \log  \frac{M}{m_W} \right) + \ldots ~, 
\end{align}
where the term in bracket is the correction to the leading-log approximation, induced by the scale-dependence of $C^{Hl}$. 
For e.g. $M = 10$ TeV, this represents a $\sim 10\%$ correction, while for $M = 10^{15}$ GeV it corresponds to a $\sim 50\%$ correction. 

In contrast, if $\gamma^i_{Hl}$ contains a term proportional to $\lambda$, the running of $\lambda$ may dominate for $C^{Hl}$ sufficiently small, because
$\beta_\lambda/\lambda > \gamma_{Hl}$.
In the seesaw, this case may occur for $C^i =C^{eH}$, see \eq{eHmW}.
Then, taking the opposite approximation of constant $C^{Hl}$ and scale-dependent $\gamma^i_{Hl}(\mu)$,
one finds that the running of $\lambda$ induces a slightly larger correction to the leading-log approximation, $\sim 30\%$ for $M = 10$ TeV and $\mathcal{O}(1)$ for $M = 10^{15}$ GeV.

%%%%%%%%%%%%%%%%%%%%%%%%%%%%%%%%%%%%%%
%%%%%%%%%%%%%%%%%%%%%%%%%%%%%%%%%%%%%%
\bibliographystyle{JHEP}
\bibliography{leptonEFT.bib}
%%%%%%%%%%%%%%%%%%%%%%%%%%%%%%%%%%%%%%

\end{document}